\documentclass[ twocolumn]{aastex62}

\usepackage{natbib}
\usepackage{amssymb, amsmath}
\usepackage{xcolor}
\usepackage{multirow}
 
\usepackage{ifthen}
\usepackage{natbib}
\usepackage{appendix}
\usepackage{etoolbox}
\usepackage[T1]{fontenc}
\usepackage{paralist}
\usepackage{booktabs}
\usepackage{multirow}
\usepackage{float}
\usepackage{url}
\usepackage{hyperref}
\usepackage{soul}

\hypersetup{linkcolor=blue,citecolor=blue,filecolor=cyan,urlcolor=blue}

\shorttitle{X-ray spectral variability of TeV blazars in the RXTE era}
\shortauthors{Wang {\em et al.}}

\begin{document}

\title{Systematic investigation of X-ray spectral variability of TeV blazars during flares in the \textit{RXTE} era}

\author{Yijun~Wang}\thanks{E-mail: yuxuan@mail.ustc.edu.cn}
\affiliation{CAS Key Laboratory for Research in Galaxies and Cosmology, Department of Astronomy, University of Science and Technology of China, Hefei 230026, People's Republic of China}
\affiliation{School of Astronomy and Space Science, University of Science and Technology of China, Hefei 230026, People's Republic of China}

\author{Yongquan~Xue}\thanks{E-mail: xuey@ustc.edu.cn}
\affiliation{CAS Key Laboratory for Research in Galaxies and Cosmology, Department of Astronomy, University of Science and Technology of China, Hefei 230026, People's Republic of China}
\affiliation{School of Astronomy and Space Science, University of Science and Technology of China, Hefei 230026, People's Republic of China}

\author{Shifu~Zhu}
\affiliation{Department of Astronomy $\&$ Astrophysics, The Pennsylvania State University, University Park, PA 16802, USA}
\affiliation{Institute for Gravitation and the Cosmos, The Pennsylvania State University, University Park, PA 16802, USA}

\author{Junhui~Fan}
\affiliation{Center for Astrophysics, Guangzhou University, Guangzhou 510006, People's Republic of China}
\affiliation{Astronomy Science and Technology Research Laboratory of Department of Education of Guangdong Province, Guangzhou 510006, People's Republic of China}

\begin{abstract}

Utilizing all the 16-year \textit{RXTE} observations, we analyze the X-ray spectra of 32 TeV blazars, and perform a systematic investigation of X-ray spectral variability for the 5 brightest sources during their major flares that lasted several days. We obtain photon spectral index ($\alpha$), flux and synchrotron radiation peak energy ($E_{\rm{p}}$) from empirical spectral fitting, and electron spectral index ($p$) from theoretical synchrotron radiation modeling. We find that both $\alpha$ and $p$ generally display a harder-when-brighter trend, confirming the results of many previous works. Furthermore, we confirm and strengthen the result that $p$ must vary in order to explain the observed X-ray spectral variability during flares, which would have useful implications for interpreting the associated higher-energy spectral variability. We see apparent electron spectral hysteresis in many but not all $p$-flux plots that takes a form of ``loop'' or oblique ``8''. We obtain a tight $p$-hardness ratio (HR) relation and a tighter $p$-$\alpha$ relation using spectra of flaring periods, both of which are also applicable to stacked data of quiescent periods. We demonstrate that these two empirical relations can be used efficiently to estimate $p$ from HR or $\alpha$ that is readily achieved. Finally, we find that, when considering TeV blazars as a whole, $\alpha$ and X-ray luminosity are positively correlated, $E_{\rm{p}}$ is negatively correlated with $p$ and $\alpha$, and $E_{\rm{p}}$ is positively correlated with HR; all these correlations are in line with the blazar sequence. However, after correcting for the Doppler boosting effect, $\alpha$ and intrinsic X-ray luminosity follow an anti-correlation.

\end{abstract}


\section{Introduction}
\label{introd}

Blazars, including Flat Spectrum Radio Quasars (FSRQs) and BL Lac objects, are a subclass of active galactic nuclei (AGNs), with one of their relativistic jets pointing to the observers at small angles \citep{1995PASP..107..803U}. They are the most important contributor to the cosmic TeV background radiation among extragalactic sources \citep[e.g.,][]{2014BrJPh..44..450H, 2015AASP....5...59S, 2016ApJ...818..187I}. Their jet emission presents double-hump broadband spectral energy distributions (SEDs) and intense variability in multiple wavelengths across different timescales. The low-energy peak is located between infrared and X-ray energies, and the high-energy peak is located at hard X-ray up to TeV $\gamma$-ray emission \citep[e.g.,][]{2010ApJ...716...30A}. The first hump is widely believed to be produced by the synchrotron radiation of relativistic electrons and/or positrons in the jet, while the origin of the second hump is still in dispute. In leptonic scenarios, the high-energy hump is dominated by the inverse Compton radiation derived from relativistic electrons scattering synchrotron photons \citep[e.g.,][]{1992ApJ...397L...5M,1998A&A...333..452K} and/or external photons, e.g., from the accretion disk, broad-line region or cosmic microwave background \citep[e.g.,][]{1992A&A...256L..27D,1994ApJ...421..153S}. In hadronic scenarios, the high-energy radiation is due to the proton emission processes \citep[e.g.,][]{2000NewA....5..377A,2001APh....15..121M,2003ApJ...586...79A,2003APh....18..593M,2013ApJ...768...54B,2015APh....70...54F}. 

According to the peak frequency of the low-energy hump ($\nu_\textrm{p}$), BL Lac objects can be divided into high-energy peaked BL Lac objects (HBLs), intermediate-energy peaked BL Lac objects (IBLs) and low-energy peaked BL Lac objects \citep[LBLs; e.g.,][]{1995ApJ...444..567P, 1998MNRAS.299..433F, 2010ApJ...716...30A,2016ApJS..226...20F}. For HBLs, the synchrotron peak is located in the UV to X-ray domain ($\nu_{\textrm{p}} > 10^{15}\ \textrm{Hz}$); for IBLs, the peak is between optical and UV regimes ($10^{14} < \nu_\textrm{p} \le 10^{15}\ \textrm{Hz}$); for LBLs, the peak is in the infrared band \citep[$\nu_\textrm{p} \le 10^{14}\ \textrm{Hz}$;][]{2010ApJ...716...30A}. Another type of blazars, FSRQs, are the high-luminosity sources whose synchrotron peak is located in the broad regime from the far-infrared to optical and even to UV wavelengths, and whose X-ray emission is from inverse Compton radiation process. As one type of sources detected at TeV energies, TeV blazars mainly belong to HBLs whose X-ray spectrum is usually dominated by synchrotron emission, therefore we use the synchrotron emission model to fit X-ray spectra of TeV blazars in this work. 

The intense variability of blazars has been illustrated by, e.g., their multiple discrete X-ray flares at timescales from several months to days to minutes \citep[e.g.,][]{2004ApJ...605..662C,2005ApJ...622..160X}, with some extremely rapid flares even having characteristic rising timescales down to half a minute \citep{2018ApJ...853...34Z}. The flaring activities are often thought to be associated with several physical processes, such as the internal shocks generated in the jet \citep{1978MNRAS.184P..61R,2001MNRAS.325.1559S}, the magnetic reconnection processes in the jet \citep{2003NewAR..47..513L,2009MNRAS.395L..29G}, or the ejection events of relativistic particles into the jet \citep{1997A&A...324..395B,1997A&A...320...19M}. Furthermore, many studies have revealed a harder-when-brighter trend in X-ray flares of blazars, which manifests itself in hardening of spectra with increasing fluxes \citep[e.g.,][]{1990ApJ...356..432G,1994ApJ...434..468S,2006ApJ...647..194X,2010ApJ...710.1271A}. \citet{2006ApJ...647..194X} used the synchrotron model to investigate the X-ray spectral variability of Mkr~421 and Mrk~501 during flares that lasted for several days. Among the four key parameters (particle spectral index, maximum Lorentz factor, total energy density and magnetic field), they found that the electron spectral index ($p$) must vary during the flaring period and it tends to decrease with increasing flux. Therefore, studying the evolution of physical parameters during flares could help us understand the underlying physical mechanism in the flaring process.

In this paper, we make use of all the 16-year archival data of TeV blazars from Proportional Counter Array (PCA) onboard \textit{Rossi X-Ray Timing Explorer} (\textit{RXTE}), a synchrotron radiation model, and the Markov chain Monte Carlo (MCMC) method to carry out a systematic investigation of the 3--25 keV X-ray spectral variability during flares of Mrk~421, Mrk~501, PKS~2155--304, PKS~2005--489, and 1ES~1959+650. This work builds on and extends further the work of \citet{2006ApJ...647..194X}, and some significant improvements over \citet{2006ApJ...647..194X} are: our target sources for detailed analysis increase from two (i.e., Mrk~421 and Mrk~501) to five; the search of target flares covers all the \textit{RXTE}/PCA observations throughout its entire lifespan ($\sim$16~years); we utilize a new method that greatly improves calculation efficiency. One primary goal of this paper is to test the universality of the conclusion in \citet{2006ApJ...647..194X} that multiple parameters (that characterize the electron distribution and magnetic field), in particular, the electron spectral index, must vary, in order to account for the observed X-ray spectral variability of TeV blazars during flares that lasted for days to weeks. One thing worth noting is that we only focus on the evolution of electron spectral index in this paper, as the other parameters are generally constrained poorly \citep[see the detailed discussion in][]{2006ApJ...647..194X}. 

\begin{center}
\begin{deluxetable*}{lcccccccc}
\tabletypesize{\footnotesize}
\tablecaption{Summary of 32 TeV blazars observed by \textit{RXTE}/PCA \label{table:allsources}}
\setlength{\tabcolsep}{3pt}

\tablehead{\colhead{Object\ \ \ \ \ \ \ \ \ \ \ \ \ \ \ \ \ \ \ \ \ \ }                          &
           \colhead{Type}              &
           \colhead{Redshift(\emph{z})}                  &
           \colhead{$\textit{N}_{\textrm{H,Gal}}^{\textrm{(DL)}}$}         &
           \colhead{$\textit{N}_\textrm{spectra}$}                  &    
           \colhead{$\chi^\textrm{2}_{\nu\textrm{,95\%}}$(CPL)}               &
           \colhead{$\textit{F}_{\textrm{max,3--25 keV}}$(CPL)}    &
           \colhead{$\chi^\textrm{2}_{\nu\textrm{,95\%}}$(LP)}   &
           \colhead{$\textit{F}_{\textrm{max,3--25 keV}}$(LP)}   \\
           \colhead{}   &
           \colhead{}   &
           \colhead{}  &
           \colhead{$\textrm{[10}^\textrm{20}\ \textrm{cm}^{\textrm{--2}}\textrm{]}$}  &
           \colhead{}   &
           \colhead{}   &
           \colhead{$\textrm{[} \textrm{10}^\textrm{--12}\ \textrm{erg}\ \textrm{cm}^{\textrm{--2}}\ \textrm{s}^{\textrm{--1}}\textrm{]}$}   &
           \colhead{}         &
           \colhead{$\textrm{[} \textrm{10}^\textrm{--12}\ \textrm{erg}\ \textrm{cm}^{\textrm{--2}}\ \textrm{s}^{\textrm{--1}}\textrm{]}$}     \\
           \colhead{(1)\ \ \ \ \ \ \ \ \ \ \ \ \ \ \ \ \ \ \ \ \ \ \ }        &
           \colhead{(2)}        &
           \colhead{(3)}        &
           \colhead{(4)}        &
           \colhead{(5)}        &
           \colhead{(6)}        &
           \colhead{(7)}        &
           \colhead{(8)}        &
           \colhead{(9)}        }                                  
\startdata
AP~Librae       & LBL  & 0.049 & 8.76 & 4    & 0.7035 & 8.18 & 0.7027 & 9.20  \\
3C~66A          & IBL  & \nodata & 8.99 & 99      & 0.8183 & 12.48 & 0.8198 & 13.41  \\
BL~Lacertae     & IBL  & 0.069 & 21.3 & 1382 & 0.7649 & 68.04 & 0.7559 & 68.38  \\
MAGIC~J2001+435 & IBL  & \nodata & 47.4 & 23      & 0.8488 & 4.96 & 0.8488 & 5.23  \\
S5~0716+714     & IBL  & 0.310 & 3.81 & 230  & 0.7965 & 19.67 & 0.7865 & 19.06  \\
W~Coma          & IBL  & 0.102 & 1.88 & 13   & 0.7827 & 3.16 & 0.7826 & 4.10  \\
1ES~0229+200    & HBL  & 0.140 & 9.21 & 205  & 0.7519 & 56.27 & 0.7469 & 57.33  \\
1ES~0414+009    & HBL  & 0.287 & 10.3 & 13   & 0.7377 & 23.41 & 0.7233 & 23.63  \\
1ES~0647+250    & HBL  & 0.450 & 12.8 & 20   & 0.7475 & 34.11 & 0.7484 & 35.08  \\
1ES~0806+524    & HBL  & 0.138 & 4.43 & 20   & 0.7226 & 7.37 & 0.7262 & 7.71  \\
1ES~1101--232    & HBL  & 0.186 & 5.76 & 99   & 0.7709 & 57.13 & 0.7722 & 59.88  \\
1ES~1215+303    & HBL  & 0.130 & 1.69 & 2    & 0.8500 & 24.93 & 0.8439 & 25.25  \\
1ES~1218+304    & HBL  & 0.182 & 1.73 & 23   & 0.7640 & 187.23 & 0.7623 & 191.27  \\
1ES~1727+502    & HBL  & 0.055 & 2.75 & 17   & 0.6961 & 29.58 & 0.7214 & 30.00  \\
1ES~1741+196    & HBL  & 0.084 & 6.86 & 12   & 0.8649 & 27.92 & 0.8546 & 29.37  \\
1ES~1959+650$^{\bigstar}$ & HBL & 0.048 & 10.1 & 146 & 0.8966 & 854.55 & 0.8558 & 862.13  \\
1ES~2344+514    & HBL  & 0.044 & 16.3 & 53   & 0.8733 & 157.59 & 0.8724 & 160.70  \\
H~1426+428      & HBL  & 0.129 & 1.38 & 165  & 0.7850 & 102.78 & 0.7781 & 104.46  \\
H~2356--309      & HBL  & 0.165 & 1.33 & 2    & 0.5959 & 8.08 & 0.5988 & 8.74  \\
Markarian~180   & HBL  & 0.045 & 1.42 & 13   & 0.6644 & 17.74 & 0.6636 & 18.55  \\
Markarian~421$^{\bigstar}$ & HBL & 0.031 & 1.38 & 1195 & 0.9335 & 3208.60 & 0.9010 & 3209.60  \\
Markarian~501$^{\bigstar}$ & HBL & 0.034 & 1.71 & 495 & 0.8771 & 1003.30 & 0.8747 & 1003.80  \\
PG~1553+113     & HBL  & 0.500 & 3.67 & 48   & 0.7529 & 28.57 & 0.7461 & 29.87  \\
PKS~0447--439    & HBL  & 0.343 & 1.78 & 11   & 0.8627 & 15.88 & 0.8657 & 17.70  \\
PKS~0548--322    & HBL  & 0.069 & 2.19 & 5    & 0.5474 & 40.76 & 0.5438 & 42.25  \\
PKS~1424+240    & HBL  & \nodata & 2.64 & 64   & 0.8207 & 3.38 & 0.8209 & 3.45   \\
PKS~2005--489$^{\bigstar}$ & HBL & 0.071 & 5.08 & 161 & 0.8221 & 332.77 & 0.8150 & 333.69  \\
PKS~2155--304$^{\bigstar}$ & HBL & 0.116 & 1.69 & 502 & 0.8111 & 207.60 & 0.8110 & 213.76  \\
RGB~J0152+017   & HBL  & 0.080 & 2.86 & 22   & 0.6224 & 9.02 & 0.6482 & 9.61  \\
RGB~J0710+591   & HBL  & 0.125 & 5.60 & 10   & 0.7979 & 64.71 & 0.7942 & 65.88  \\
3C~279          & FSRQ & 0.536 & 2.21 & 1979 & 0.7548 & 75.37 & 0.7483 & 75.85  \\
PKS~1510--089    & FSRQ & 0.361 & 7.96 & 1314 & 0.7526 & 45.80 & 0.7382 & 48.54  \\
\enddata
\tablenotetext{}{
\textbf{N\textsc{ote.}} (1) Object name (objects marked with a $\bigstar$ sign represent the sources selected for further spectral fitting with the synchrotron radiation model). (2) Source type, provided by \href{http://tevcat.uchicago.edu/}{TeVCat}. (3) Redshift, provided by \href{http://tevcat.uchicago.edu/}{TeVCat}. (4) Galactic hydrogen column density along the line of sight, provided by DL (Dickey \& Lockman 1990). (5) Number of observations performed by \textit{RXTE}/PCA during its 16-year lifespan. 
(6) Reduced $\chi^\textrm{2}$ value that is larger than 95\% of the best-fit $\chi^\textrm{2}_\nu$ values when fitting spectra with a cut-off power-law (CPL) model. 
(7) Maximum 3--25 keV flux obtained by fitting spectra with CPL (16-year data). 
(8) Reduced $\chi^\textrm{2}$ value that is larger than 95\% of the best-fit $\chi^\textrm{2}_\nu$ values when fitting spectra with a log-parabolic (LP) model.
(9) Maximum 3--25 keV flux obtained by fitting spectra with LP (16-year data).}
\end{deluxetable*}
\end{center}

\section{DATA AND DATA REDUCTION}
\label{sec:obser}

\textit{RXTE}, carrying All Sky Monitor (ASM), PCA, and High-Energy X-Ray Timing Experiment (HEXTE), started operation in 1996 January and completed its scientific mission in 2012 January. During its 16-year lifespan, \textit{RXTE} had observed 52 blazars \citep{2013ApJ...772..114R}, including 32 TeV blazars (see Table~\ref{table:allsources}) verified in the catalog of TeV sources (i.e., TeVCat\footnote{{The TeVCat online catalog is provided} by Scott Wakely \& Deirdre Horan (\href{http://tevcat.uchicago.edu/}{http://tevcat.uchicago.edu/}).}). 
These TeV blazars are 2~FSRQs, 1~LBLs, 5~IBLs, and 24~HBLs.

\begin{center}
\begin{deluxetable*}{lcccccccccccc}
\tabletypesize{\scriptsize}
\tablecaption{Spectral fitting results of 5 TeV blazars during flaring periods and quiescent periods  \label{table:allta}}
\setlength{\tabcolsep}{2pt}   

\tablehead{\colhead{} & \colhead{} & \colhead{} & \multicolumn{3}{c}{Cut-off Power Law} & \multicolumn{5}{c}{Log-parabolic} & \multicolumn{2}{c}{Synchrotron}  \\
           \cmidrule(l{3pt}r{3pt}){4-6} \cmidrule(l{3pt}r{3pt}){7-11} \cmidrule(l{3pt}r{3pt}){12-13}
           \colhead{Object \ \ \ \ \ \ \ \ \ \ \ \ } & \colhead{Date} & \colhead{MJD} & \colhead{$\Gamma$} & \colhead{$\textit{F}_\textrm{3--25 keV}$} &      
           \colhead{$\chi^2_\nu$ ($\nu$)} & \colhead{\textit{a}} & \colhead{\textit{b}} & 
           \colhead{\textit{K}} & \colhead{$\textit{F}_\textrm{3--25 keV}$} & 
           \colhead{$\chi^2_\nu$ ($\nu$)} &
           \colhead{$p$} & \colhead{$\chi^2_\nu$ ($\nu$)}   \\
           \colhead{(1)\ \ \ \ \ \ \ \ \ \ \ \ \ }        &
           \colhead{(2)}        &
           \colhead{(3)}        &
           \colhead{(4)}        &
           \colhead{(5)}        &
           \colhead{(6)}        &
           \colhead{(7)}        &
           \colhead{(8)}        &
           \colhead{(9)}        &                                 
           \colhead{(10)}        &
           \colhead{(11)}        &
           \colhead{(12)}        &
           \colhead{(13)}        }                                  
\startdata
Mrk~421 & $\blacktriangle$ 2001/03/20 &  51988.44 & 2.45 (0.04) &  4.52 & 0.34 (26) & 2.22 (0.10) & 0.28 (0.06) & 0.36 (0.03) &  4.53 & 0.22 (26) & $\textrm{4.08}^{+\textrm{0.05}}_{-\textrm{0.07}}$ & 0.42 (24) \\
        & $\blacktriangle$ 2001/03/20 &  51988.73 & 2.23 (0.04) &  7.49 & 0.53 (32) & 1.97 (0.08) & 0.29 (0.05) & 0.37 (0.03) &  7.50 & 0.33 (32) & $\textrm{3.58}^{+\textrm{0.07}}_{-\textrm{0.04}}$ & 0.63 (30) \\
(flaring)        & $\blacktriangle$ 2001/03/21 &  51989.69 & 2.16 (0.02) & 15.47 & 0.66 (41) & 1.80 (0.06) & 0.38 (0.03) & 0.65 (0.04) & 15.47 & 0.29 (41) & $\textrm{3.45}^{+\textrm{0.05}}_{-\textrm{0.03}}$ & 0.80 (40) \\
        & $\blacktriangle$ 2001/03/22 &  51990.04 & 2.26 (0.03) &  7.58 & 0.38 (31) & 1.97 (0.08) & 0.33 (0.05) & 0.41 (0.03) &  7.60 & 0.22 (31) & $\textrm{3.64}^{+\textrm{0.06}}_{-\textrm{0.05}}$ & 0.54 (28) \\
        & $\blacktriangle$ 2001/03/22 &  51990.16 & 2.17 (0.07) &  5.65 & 0.57 (27) & 1.77 (0.15) & 0.51 (0.09) & 0.28 (0.04) &  5.68 & 0.56 (27) & $\textrm{3.58}^{+\textrm{0.12}}_{-\textrm{0.13}}$ & 0.58 (21) \\
\cline{2-13}
(quiescent)        & $\blacktriangledown\ \ \ \textrm{2003}^\textrm{/04/02}_\textrm{/05/02}$ &  $\textrm{52700}^\textrm{+31}_\textrm{+63}$ & 2.51 (0.06) &  1.14 & 0.40 (22) & 2.23 (0.11) & 0.35 (0.07) & 0.10 (0.01) &  1.14 & 0.37 (22) & $\textrm{4.25}^{+\textrm{0.08}}_{-\textrm{0.08}}$ & 0.45 (21) \\
(quiescent)        & $\blacktriangledown\ \ \ \textrm{2010}^\textrm{/04/09}_\textrm{/04/18}$ &  $\textrm{55000}^\textrm{+295}_\textrm{+304}$ & 2.48 (0.10) &  0.85 & 0.56 (17) & 2.11 (0.20) & 0.49 (0.13) & 0.08 (0.01) &  0.85 & 0.47 (17) & $\textrm{4.26}^{+\textrm{0.16}}_{-\textrm{0.16}}$ & 0.64 (16) \\
\hline
Mrk~501 & $\blacktriangle$ 1997/04/12 &  50550.19 & 1.72 (0.05) &  5.25 & 0.78 (28) & 1.61 (0.10) & 0.12 (0.06) & 0.09 (0.01) &  5.25 & 0.67 (28) & $\textrm{2.43}^{+\textrm{0.11}}_{-\textrm{0.08}}$ & 0.83 (29) \\
        & $\blacktriangle$ 1997/04/12 &  50550.45 & 1.71 (0.04) &  6.00 & 0.73 (29) & 1.62 (0.10) & 0.09 (0.06) & 0.10 (0.01) &  6.00 & 0.64 (29) & $\textrm{2.39}^{+\textrm{0.09}}_{-\textrm{0.08}}$ & 0.66 (30) \\
(flaring)        & $\blacktriangle$ 1997/04/13 &  50551.46 & 1.59 (0.04) &  7.84 & 0.77 (35) & 1.46 (0.08) & 0.15 (0.05) & 0.10 (0.01) &  7.84 & 0.65 (35) & $\textrm{2.11}^{+\textrm{0.11}}_{-\textrm{0.07}}$ & 0.80 (37) \\
        & $\blacktriangle$ 1997/04/14 &  50552.34 & 1.64 (0.04) &  6.20 & 1.14 (28) & 1.52 (0.10) & 0.13 (0.06) & 0.09 (0.01) &  6.21 & 1.02 (28) & $\textrm{2.28}^{+\textrm{0.07}}_{-\textrm{0.12}}$ & 1.18 (31) \\
        & $\blacktriangle$ 1997/04/15 &  50553.27 & 1.69 (0.04) &  6.12 & 0.52 (27) & 1.58 (0.08) & 0.11 (0.05) & 0.09 (0.01) &  6.12 & 0.49 (27) & $\textrm{2.37}^{+\textrm{0.07}}_{-\textrm{0.09}}$ & 0.55 (27) \\
\cline{2-13}
   (quiescent)     & $\blacktriangledown\ \ \ \textrm{2004}^\textrm{/06/14}_\textrm{/06/21}$ & $\textrm{53100}^\textrm{+70}_\textrm{+77}$  & 2.07 (0.07) &  1.06 & 0.65 (20) & 1.82 (0.13) & 0.30 (0.08) & 0.04 (0.005) &  1.06 & 0.58 (20) & $\textrm{3.23}^{+\textrm{0.16}}_{-\textrm{0.16}}$ & 0.70 (19) \\
   (quiescent)     & $\blacktriangledown\ \ \ \textrm{1998}^\textrm{/06/19}_\textrm{/06/23}$ &  $\textrm{50980}^\textrm{+2}_\textrm{+7}$ & 2.11 (0.06) &  0.71 & 0.40 (22) & 2.04 (0.12) & 0.08 (0.07) & 0.03 (0.003) &  0.72 & 0.40 (22) & $\textrm{3.26}^{+\textrm{0.07}}_{-\textrm{0.10}}$ & 0.42 (21) \\
\hline
PKS~2155--304 & $\blacktriangle$ 1996/05/19 &  50222.65 & 2.32 (0.07) &  1.10 & 0.47 (25) & 2.16 (0.15) &  0.22 (0.10) & 0.07 (0.01) &  1.10 & 0.43 (25) & $\textrm{3.78}^{+\textrm{0.09}}_{-\textrm{0.11}}$ & 0.45 (22) \\
             & $\blacktriangle$ 1996/05/19 &  50222.80 & 2.28 (0.07) &  1.36 & 0.35 (26) & 2.15 (0.14) &  0.17 (0.09) & 0.08 (0.01) &  1.37 & 0.35 (26) & $\textrm{3.65}^{+\textrm{0.09}}_{-\textrm{0.09}}$ & 0.37 (23) \\
     (flaring)        & $\blacktriangle$ 1996/05/20 &  50223.55 & 2.25 (0.06) &  1.84 & 0.56 (28) & 2.05 (0.14) &  0.26 (0.09) & 0.10 (0.01) &  1.85 & 0.50 (28) & $\textrm{3.62}^{+\textrm{0.11}}_{-\textrm{0.09}}$ & 0.61 (25) \\
             & $\blacktriangle$ 1996/05/21 &  50223.93 & 2.21 (0.05) &  1.71 & 0.22 (26) & 1.99 (0.12) &  0.28 (0.07) & 0.09 (0.01) &  1.72 & 0.23 (26) & $\textrm{3.54}^{+\textrm{0.09}}_{-\textrm{0.08}}$ & 0.20 (23) \\
             & $\blacktriangle$ 1996/05/21 &  50224.21 & 2.32 (0.12) &  1.53 & 0.46 (23) & 2.21 (0.24) &  0.20 (0.16) & 0.11 (0.02) &  1.55 & 0.45 (23) & $\textrm{3.81}^{+\textrm{0.17}}_{-\textrm{0.17}}$ & 0.48 (21) \\
\hline
PKS~2005--489 & $\blacktriangle$ 1998/10/22 &  51108.51 & 2.17 (0.08) &  1.42 & 0.45 (23) & 1.97 (0.17) & 0.25 (0.11) & 0.07 (0.01) &  1.43 & 0.39 (23) & $\textrm{3.45}^{+\textrm{0.11}}_{-\textrm{0.17}}$ & 0.45 (19) \\
             & $\blacktriangle$ 1998/11/04 &  51121.71 & 2.03 (0.06) &  2.67 & 0.60 (26) & 1.82 (0.12) & 0.25 (0.07) & 0.09 (0.01) &  2.68 & 0.55 (26) & $\textrm{3.13}^{+\textrm{0.10}}_{-\textrm{0.11}}$ & 0.47 (23) \\
     (flaring)        & $\blacktriangle$ 1998/11/10 &  51127.63 & 2.13 (0.05) &  3.33 & 0.51 (28) & 2.01 (0.10) & 0.15 (0.06) & 0.14 (0.01) &  3.34 & 0.42 (28) & $\textrm{3.34}^{+\textrm{0.05}}_{-\textrm{0.08}}$ & 0.55 (25) \\
             & $\blacktriangle$ 1998/11/16 &  51133.49 & 2.14 (0.05) &  2.74 & 0.73 (31) & 1.97 (0.11) & 0.20 (0.07) & 0.12 (0.01) &  2.75 & 0.65 (31) & $\textrm{3.34}^{+\textrm{0.09}}_{-\textrm{0.06}}$ & 0.75 (28) \\
             & $\blacktriangle$ 1998/11/28 &  51145.43 & 2.42 (0.08) &  1.49 & 0.62 (27) & 2.25 (0.16) & 0.22 (0.10) & 0.11 (0.02) &  1.49 & 0.58 (27) & $\textrm{3.96}^{+\textrm{0.09}}_{-\textrm{0.13}}$ & 0.58 (24) \\
\cline{2-13}
       (quiescent)      & $\blacktriangledown\ \ \ \textrm{2009}^\textrm{/05/24}_\textrm{/06/03}$ &  $\textrm{54900}^\textrm{+75}_\textrm{+85}$ & 2.22 (0.05) &  1.32 & 0.62 (21) & 1.94 (0.11) & 0.33 (0.07) & 0.07 (0.006) &  1.32 & 0.47 (21) & $\textrm{3.56}^{+\textrm{0.10}}_{-\textrm{0.08}}$ & 0.72 (20) \\
\hline
1ES~1959+650 & $\blacktriangle$ 2002/05/19 &  52413.00 & 1.92 (0.13) &  1.46 & 0.92 (21) & 1.83 (0.26) & 0.13 (0.16) & 0.09 (0.02) &  1.46 & 0.94 (21) & $\textrm{2.83}^{+\textrm{0.26}}_{-\textrm{0.18}}$ & 1.03 (17) \\
             & $\blacktriangle$ 2002/05/19 &  52413.78 & 1.56 (0.05) &  1.96 & 0.86 (26) & 1.26 (0.11) & 0.33 (0.07) & 0.08 (0.01) &  1.96 & 0.78 (26) & $\textrm{1.99}^{+\textrm{0.13}}_{-\textrm{0.15}}$ & 0.93 (24) \\
  (flaring)           & $\blacktriangle$ 2002/05/20 &  52414.47 & 1.57 (0.04) &  2.02 & 1.01 (31) & 1.36 (0.09) & 0.22 (0.05) & 0.10 (0.01) &  2.03 & 0.97 (31) & $\textrm{2.06}^{+\textrm{0.13}}_{-\textrm{0.12}}$ & 1.01 (27) \\
             & $\blacktriangle$ 2002/05/21 &  52415.14 & 1.66 (0.13) &  2.07 & 1.03 (25) & 1.30 (0.26) & 0.40 (0.16) & 0.06 (0.01) &  2.08 & 0.99 (25) & $\textrm{2.02}^{+\textrm{0.47}}_{-\textrm{0.27}}$ & 1.19 (21) \\
             & $\blacktriangle$ 2002/05/21 &  52415.39 & 1.80 (0.08) &  2.23 & 0.83 (23) & 1.59 (0.16) & 0.22 (0.10) & 0.07 (0.01) &  2.24 & 0.78 (23) & $\textrm{2.54}^{+\textrm{0.18}}_{-\textrm{0.14}}$ & 0.93 (21) \\
\cline{2-13}
  (quiescent)           & $\blacktriangledown\ \ \ \textrm{2003}^\textrm{/05/24}_\textrm{/06/07}$ &  $\textrm{52700}^\textrm{+83}_\textrm{+97}$ & 2.30 (0.09) &  0.84 & 0.60 (22) & 2.11 (0.17) & 0.30 (0.11) & 0.06 (0.008) &  0.85 & 0.57 (22) & $\textrm{3.85}^{+\textrm{0.09}}_{-\textrm{0.18}}$ & 0.61 (21) \\

\enddata
\tablenotetext{}{
\textbf{N\textsc{ote.}} (1) Object name.
(2, 3) Date and Modified Julian date (MJD) of the \textit{RXTE}/PCA observation. 
(4--6) Photon index ($\Gamma$; $\Gamma$=$\alpha$+1 and $\alpha$ is the photon spectral index),
3--25~keV flux (in the unit of $\textrm{10}^\textrm{--10}\ \textrm{erg}\ \textrm{cm}^{\textrm{--2}}\ \textrm{s}^{\textrm{--1}}$), and
best-fit reduced chi-square ($\chi^\textrm{2}_\nu$; degree of freedom $\nu$) with cut-off power-law fitting, respectively. 
(7--11) Parameter \textit{a}, \textit{b}, and \textit{K} (in the unit of photon $\textrm{cm}^{\textrm{--2}}\ \textrm{s}^{\textrm{--1}}\ \textrm{keV}^{\textrm{--1}}$) in Eq.~(\ref{equ:lppp}), 
3--25 keV flux (in the unit of $\textrm{10}^\textrm{--10}\ \textrm{erg}\ \textrm{cm}^{\textrm{--2}}\ \textrm{s}^{\textrm{--1}}$), and
best-fit $\chi^\textrm{2}_\nu$ with log-parabolic fitting, respectively.
(12, 13) Electron spectral index ($p$) and best-fit $\chi^\textrm{2}_\nu$ with synchrotron model fitting (see Section~\ref{sec:model}), respectively. 
All errors represent 1-$\sigma$ errors. 
For each source, 
the rows marked with ``$\blacktriangle$'' represent the results of 5 observations occurring during an example flare as plotted in Figure~\ref{fig:spectrum} (i.e., the 5 colourfully-encircled data points in the light curves), and
the rows marked with ``$\blacktriangledown$'' represent the results of stacked spectra during quiescent period (the date and MJD indicate the time range of stacked quiescent-period observations).}
\end{deluxetable*}
\end{center}

\subsection{All Data}
\label{sec:tdata}
In this paper, we utilized data from PCA that consists of five nearly identical proportional counter units (PCUs). 
For the 16-year observations of the 32 TeV blazars, we followed \citet{2011ApJS..193....3R} to extract the first xenon layer data from PCU~0, PCU~1, and PCU~2 before 1998 December 23; PCU~0 and PCU~2 between 1998 December 23 and 2000 May 12; and PCU~2 after 2000 May 12, respectively, given that PCUs 1, 3 and 4 suffered from high-voltage break-down issues during their on-source time, and PCU 0 had been operating without its propane layer since 2000 May 12. In this work, we made use of Standard2 data exclusively, and binned each individual PCA observation into one data point when producing light curves (the median exposure time of all observations of each source is more than 1000 s). The numbers of PCA observations of these objects are summarized in Table~\ref{table:allsources} (column~5).  

We followed \citet{2005ApJ...622..160X} and \citet{2006ApJ...647..194X} to perform data reduction and analysis using {\sc{ftools}} version 6.19. Firstly, we created the data filter file and good time intervals (GTIs) file for each observation following the standard procedure\footnote{\href{http://heasarc.gsfc.nasa.gov/docs/xte/recipes/cook\_book.html}{http://heasarc.gsfc.nasa.gov/docs/xte/recipes/cook\_book.html}.}. Secondly, according to the suggested criterion\footnote{{Details can be found in the part of} ``Important Downloads and Links'' at \href{http://heasarc.gsfc.nasa.gov/docs/xte/pca\_news.html}{http://heasarc.gsfc.nasa.gov/docs/xte/pca\_news.html}.}, we used the latest faint background model (pca\_bkgd\_cmfaintl7\_eMv20051128.mdl) for observations with count rates $<$ 40 c/s/PCU and bright background model (pca\_bkgd\_cmbrightvle\_eMv20051128.mdl) for observations with count rates $\ge$ 40 c/s/PCU to simulate background events. Finally, we extracted spectra for both observational data and simulated background events using corresponding GTIs, and grouped the spectra appropriately using {\sc{grppha}} in order to improve the signal-to-noise ratio (S/N) for subsequent spectral analysis.

\subsection{Data of Flaring \textit{Periods}}
\label{sec:fdata}

Since one of the major goals of this work is to study the 3--25 keV X-ray spectral variability during flares, the subsequent analysis has been limited to objects with high X-ray fluxes and at least 5 observations during one flare.

As such, among 32 TeV blazars, we singled out 5 objects (i.e., Mrk~421, Mrk~501, PKS~2155--304, PKS~2005--489, and 1ES~1959+650; hereafter ``the five sources''; see Table~\ref{table:allta}) and for which \textit{RXTE}/PCA data allow us to obtain high-quality spectra (detailed analysis of  the other 27 TeV blazars will be presented in a future study). Furthermore, the flares were selected with the following criteria: 1) individual flares lasted for several days and were covered by at least two observations in both the rise and decay periods; and 2) the minimum total count rate (summed over available PCUs) in 3--25~keV is above 30 c/s. In addition, adjacent outbursts following or followed by those flares were also included. Finally, we picked out 20.5 flares for Mrk~421 (an outburst with observations only in the rise or decay period was considered as 0.5 flare), 7 flares for Mrk~501, 4 flares for PKS~2155--304, and only one flare for both PKS~2005--489 and 1ES~1959+650 from the 16-year data (see Figures~\ref{fig:spectrum} and \ref{fig:pfevo} for the typical flares of each source; also see the observations annotated with ``$\blacktriangle$'' in Table~\ref{table:allta}. 

\subsection{Data of Quiescent \textit{Periods}}
\label{sec:ldata}
As a comparison, we also extracted spectra for the above five objects when they stayed in the relatively quiescent periods, with variability amplitude being relatively small over several days. In view of the low S/N of each spectrum, we stacked multiple spectra within a certain time range. Finally, we produced two stacked spectra for both Mrk~421 and Mrk~501, one stacked spectrum for both PKS~2155--304 and 1ES~1959+650, and no stacked spectrum for PKS~2155--304, respectively (see the rows annotated with ``$\blacktriangledown$'' in Table~\ref{table:allta}). For PKS~2155--304,  its two stacked spectra of the quiescent period are concave and thus not included in Table \ref{table:allta}, because its 3--25 keV X-ray emission likely comes from both the synchrotron radiation and inverse Compton scattering processes, and our SED modeling of these two spectra demonstrates that the synchrotron radiation model could not constrain the value of $p$ well.

The hierarchical X-ray flaring phenomenon has been observed in multiple blazars \citep[e.g.,][]{2004ApJ...605..662C,2005ApJ...622..160X}, which indicates that flares could occur at timescales from minutes to months, and X-ray light curves manifest the superposition of these events at different timescales. Therefore, there might be no true state transition in blazars, even though their fluxes vary largely. In this work, we selected flaring periods and relatively quiescent periods at several-day timescales based on the aforementioned selection criteria and our visual inspection.

\section{SPECTRAL FITTING, MODELING, AND METHOD}
\label{sec:model}

\subsection{Photon spectral Analysis}
\label{sec:xspec}
For all the spectra of 32 TeV blazars, we performed spectral analysis with the {\sc{xspec}} software package \citep[version 12.9.0;][]{1996ASPC..101...17A}. For each spectrum, we experimented with four empirical models: power law, broken power law, power law with an exponential cut-off, and log-parabolic. 
For each object, we fixed the Galactic hydrogen absorption parameter ($\textit{N}_{\textrm{H}}$) that was from the survey by \citet{1990ARA&A..28..215D}, as reported in Table~\ref{table:allsources}. 

According to the distribution of reduced chi-square when fitting each source, we found that both the cut-off power law and log-parabolic models provided better fits to the data than power law and broken power law models. And it is often difficult to decide which is the best-fit model between  the cut-off power law and log-parabolic models (see Table~\ref{table:allsources}).  Here, we fitted the spectra with cut-off power law to obtain the photon spectral index (\textrm{$\alpha$}) (see details in Section~\ref{sec:esipsi}), and with log-parabolic model to obtain the peak energy ($E_{\rm p}$) of the synchrotron radiation hump in SED (see details in Section~\ref{sec:peakes}). 
The intrinsic SEDs (i.e., corrected for Galactic absorption) derived with the best fits were subsequently used for synchrotron radiation modeling, where we adopted the following cosmological parameters: $\textit{H}_\textrm{0}$=70 km s$^{\textrm{-1}}$ Mpc$^{\textrm{-1}}$, $\textrm{$\Omega$}_\textit{m}$=0.28, and $\textrm{$\Omega$}_{\textrm{$\Lambda$}}$=0.72 \citep{2013ApJS..208...19H}.

\subsection{Synchrotron Model}
\label{sec:theom}
We used the homogeneous synchrotron radiation model presented in \citet{2006ApJ...647..194X} to fit the time-resolved flaring-period spectra (see Section~\ref{sec:fdata}) and the stacked quiescent-period spectra (see Section~\ref{sec:ldata}) of the aforementioned five sources. It was based on the assumption that a single flare is generated from a localized region of the jet (i.e., jet blob\footnote{Synchrotron radiation models have also been successfully used to describe the emission of non-blazar jet blobs/knots \citep[e.g.,][]{2002ApJ...564..683M,2003ApJ...586L..41H,2006ApJ...640..211H}.}), where the spatial distribution of electrons and magnetic field is homogeneous.

\subsubsection{Electron spectral distribution}
\label{sec:parti}
Full details on the synchrotron radiation model were presented in Section~3 of \citet{2006ApJ...647..194X}; here we only provide a brief introduction. We assume that the emitting electrons follow the power-law spectral distribution with power-law index $p$ and low- and high-energy cutoffs, $\textrm{$\textrm{$\textrm{$\gamma$}$}$}_{\textrm{min}}$ and $\textrm{$\textrm{$\textrm{$\gamma$}$}$}_{\textrm{max}}$, and are homogeneously distributed in the emitting region. In addition, the emitting region is assumed to be a spherical zone with the radius of \textit{r} that is compatible with the duration of the flare. Then we can evaluate the spectrum of the emission by integrating the differential power of synchrotron radiation over the entire Lorentz-factor range (i.e., $\textrm{$\textrm{$\textrm{$\gamma$}$}$}_{\textrm{min}}\le \gamma\le \textrm{$\textrm{$\textrm{$\gamma$}$}$}_{\textrm{max}}$) within the jet blob.   

We chose electron spectral index ($p$), magnetic field (\textit{B}), maximum Lorentz factor of electrons ($\textrm{$\gamma$}_{\textrm{max}}$), and total energy density of electrons ($\textit{E}_{\textrm{tot}}$/$\textit{m}_\textit{e} \textit{c}^\textmd{2}$) as free parameters when performing synchrotron radiation modeling. During the fitting process, the Doppler factor (\textrm{$\delta$}) and the minimum Lorentz factor of the electrons ($\textrm{$\gamma$}_{\textrm{min}}$) were frozen, i.e., \textrm{$\delta$}=15 (a nominal value for TeV blazars) and $\textrm{$\gamma$}_{\textrm{min}}$=10$^\textrm{4}$. We had verified that reasonable change of \textrm{$\delta$} and $\textrm{$\gamma$}_{\textrm{min}}$ values had little impact on the distribution of $p$. We found that our homogeneous synchrotron radiation model with the above parameter settings can well produce the observed 3--25 keV spectra, as in \citet{2006ApJ...647..194X}.

\subsubsection{Fitting method}
\label{sec:method}
In \citet{2006ApJ...647..194X}, the statistically acceptable solutions were obtained through grid search. In this papar, we used a new method to obtain the solutions. At first, we defined the sufficiently-wide preliminary ranges of the four parameters, i.e., $p$ in 1.00--5.00, \textit{B} in $\textrm{10}^{\textrm{-3}}$--$\textrm{10}^{\textrm{2}}$ G, $\textrm{$\gamma$}_\textrm{max}$ in $\textrm{10}^{\textrm{5}}$--$\textrm{10}^{\textrm{11}}$ (note that the final solutions are selected in the realistic range of $\textrm{10}^{\textrm{5}}$--$\textrm{10}^{\textrm{8}}$), and $\textit{E}_{\textrm{tot}}$/$\textit{m}_\textit{e} \textit{c}^\textrm{2}$  cm$^{\textrm{-3}}$ in $\textrm{10}^{\textrm{3}}$--$\textrm{10}^{\textrm{10}}$ ergs cm$^{\textrm{-3}}$, respectively. We adopted linear steps for $p$, $\log {(\textit{B})}$, $\log {(\textrm{$\gamma$}_{\textrm {max}})}$, and $\log {(\textit{E}_{\textrm{tot}})}$ in grid search; the step was 0.01 for $p$ and 0.02 for the other three parameters, respectively. Starting with the preliminary parameter ranges, we used MPFIT to obtain a set of best-fit parameters that were then selected as the initial values for subsequent MCMC fitting. We utilized the MCMC method for fitting in order to narrow down the corresponding range of each parameter for each spectrum. Subsequently, we carried out a grid search to find acceptable solutions (i.e., $\chi^2_\nu \le 1+\sqrt{2/\nu}$, where $\chi^2_\nu$ is reduced chi-square and $\nu$ is degree of freedom) within the parameter ranges constrained by the MCMC method.

The acceptable solutions usually cover a small range of the entire preliminary parameter space mentioned above. The usual way of performing grid search from end to end would cover the whole parameter space uniformly, which is, however, very time-consuming. Conversely, the MCMC method could reduce the computing time to one eighth of the time needed by grid search, but the solutions might sometimes be trapped within a local minimum so that a meaningful parameter distribution cannot be obtained. Therefore, we decided to first use the MCMC method to restrict the parameter range from the preliminary range, and then used grid search to obtain the final solutions (and thus the $p$ distribution). We had verified that this fitting method would obtain the same parameter distributions as the grid search method adopted by \citet{2006ApJ...647..194X} and could greatly improve computation efficiency. 
In this paper, we focus only on the $p$ distribution that is reasonably constrained, given that the distributions of \textit{B}, $\textrm{$\gamma$}_{\textrm{max}}$, and $\textit{E}_{\textrm{tot}}$ are usually constrained poorly. As indicated by \citet{2006ApJ...647..194X}, the SED modeling suffers from serious degeneracy among the other three parameters (\textit{B}, $\textrm{$\gamma$}_{\textrm{max}}$ and $\textit{E}_{\textrm{tot}}$); our result draws the same conclusion. The 1 $\sigma$ errors of $p$ are obtained based on the range of $p$ when the chi-square ($\chi^2$) equals to one plus the minimum of $\chi^2$ (i.e., the best-fit $\chi^2$) in the plot of $\chi^2$ versus $p$.

\begin{figure*}[!thbp]
	\centering
	\includegraphics[width=0.73\paperwidth, clip]{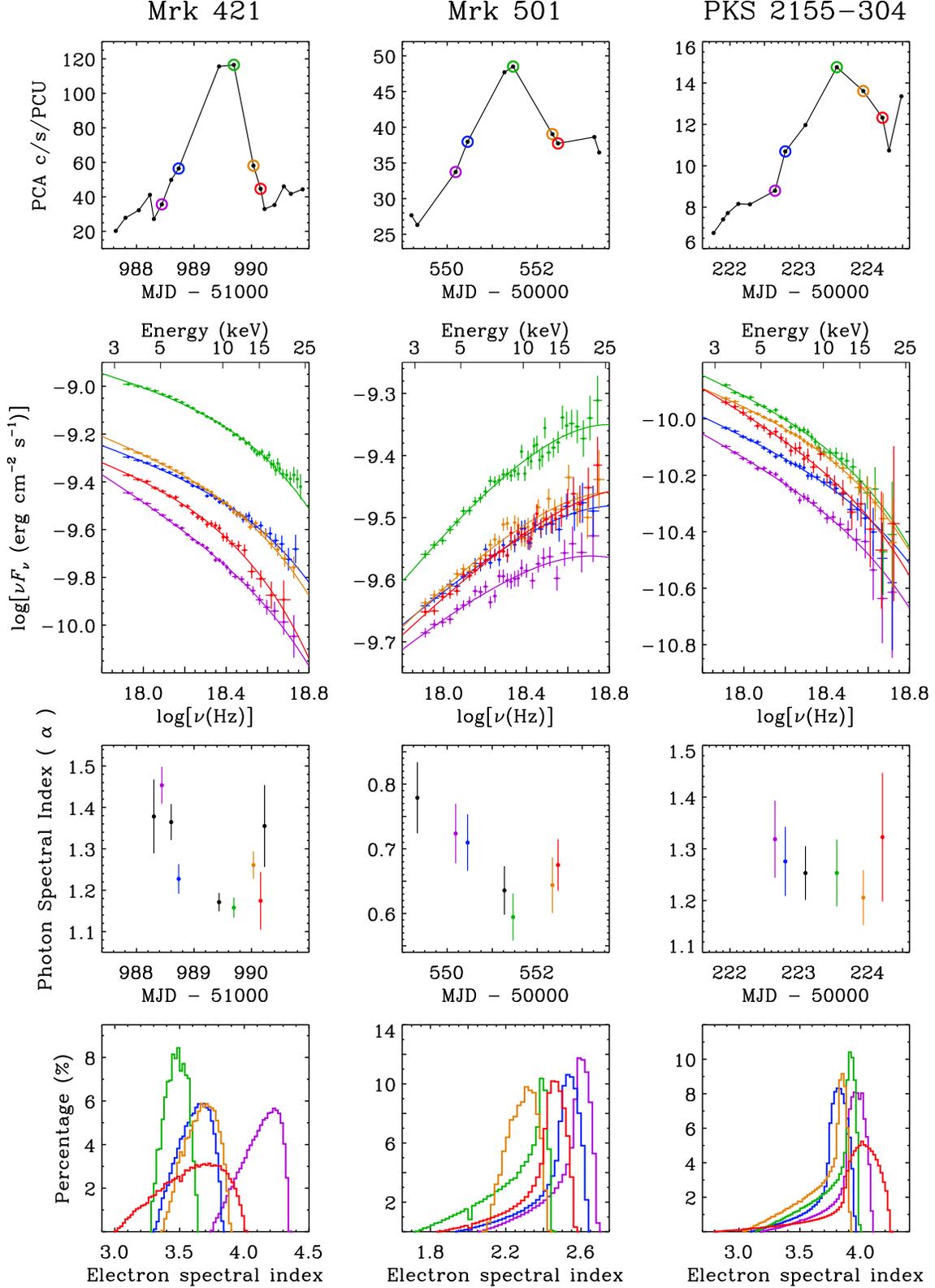}
	\caption{The 3--25 keV PCA light curves, X-ray spectra, photon spectral index variations, and normalized distributions of electron spectral index during a typical flare of Mrk~421, Mrk~501, PKS~2155--304, PKS~2005--489, and 1ES~1959+650, respectively (one column for one source).
		\textit{Top} row: 3--25 keV X-ray light curves (typical errors on count rates in units of c/s/PCU are 0.16 for Mrk~421, 0.11 for Mrk~501, 0.09 for PKS~2155-489, 0.08 for PKS~2005--489, and 0.14 for 1ES~1959+650, respectively, which are very small and therefore not plotted).
		\textit{Second} row: X-ray spectra with solid curves representing best-fit synchrotron models that were obtained using $\chi^2$ statistics (see Table~\ref{table:allsources}).
		\textit{Third} row: variations of photon spectral index ($\alpha$) over time during flares.
		\textit{Bottom} row: normalized distributions of electron spectral index ($p$) derived with the synchrotron model fitting.
		For each source (column), the same color represents the same observation. }
	\label{fig:spectrum}
\end{figure*}

\renewcommand{\thefigure}{\arabic{figure} (Cont.)}
\addtocounter{figure}{-1}

\begin{figure*}[!thbp]
	\centering
	\includegraphics[width=0.73\paperwidth, clip]{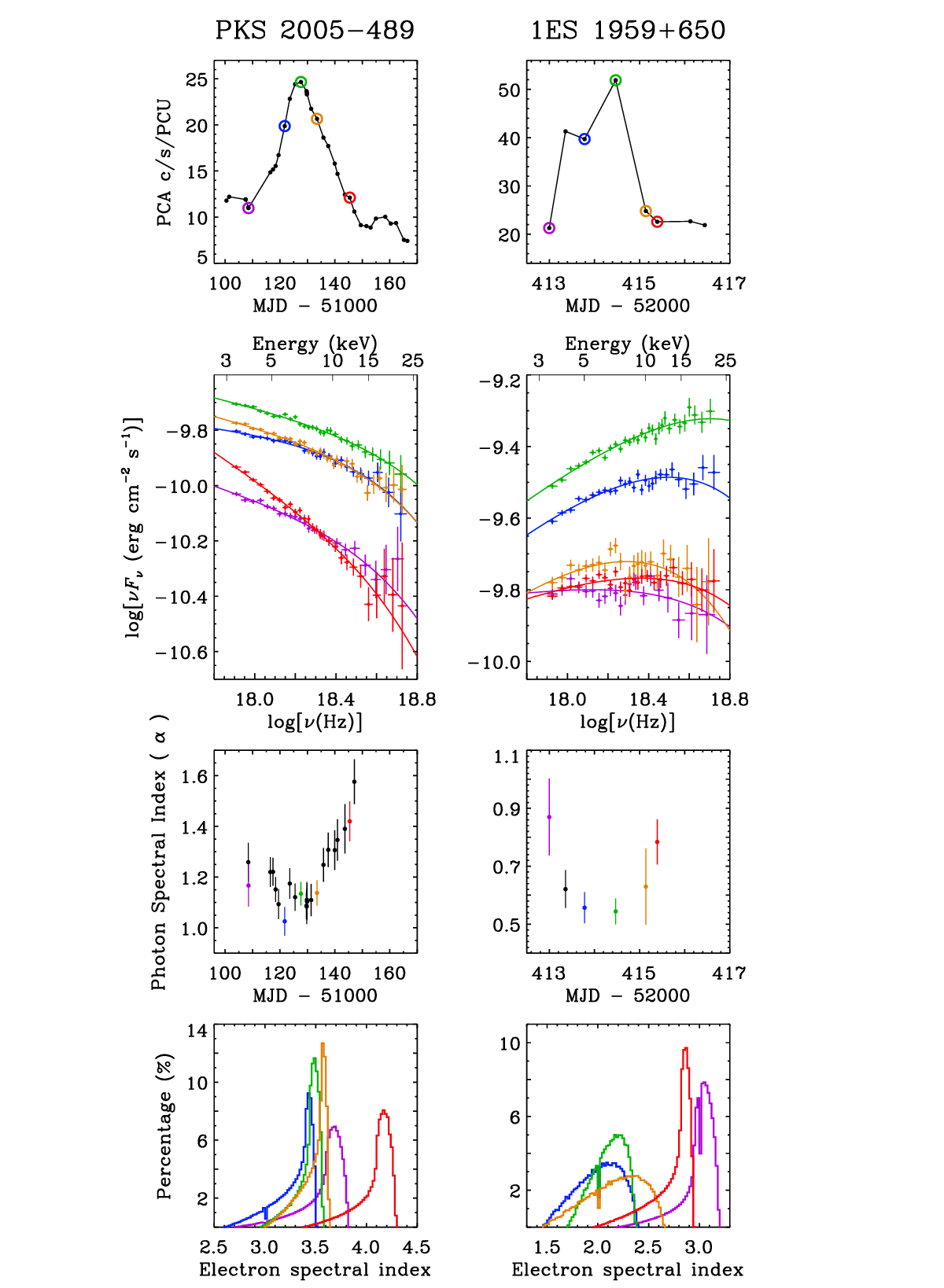}
	\caption{}
\end{figure*}

\renewcommand{\thefigure}{\arabic{figure}}

\begin{figure*}[!thbp]
	\centering
	\includegraphics[width=0.76\paperwidth, clip]{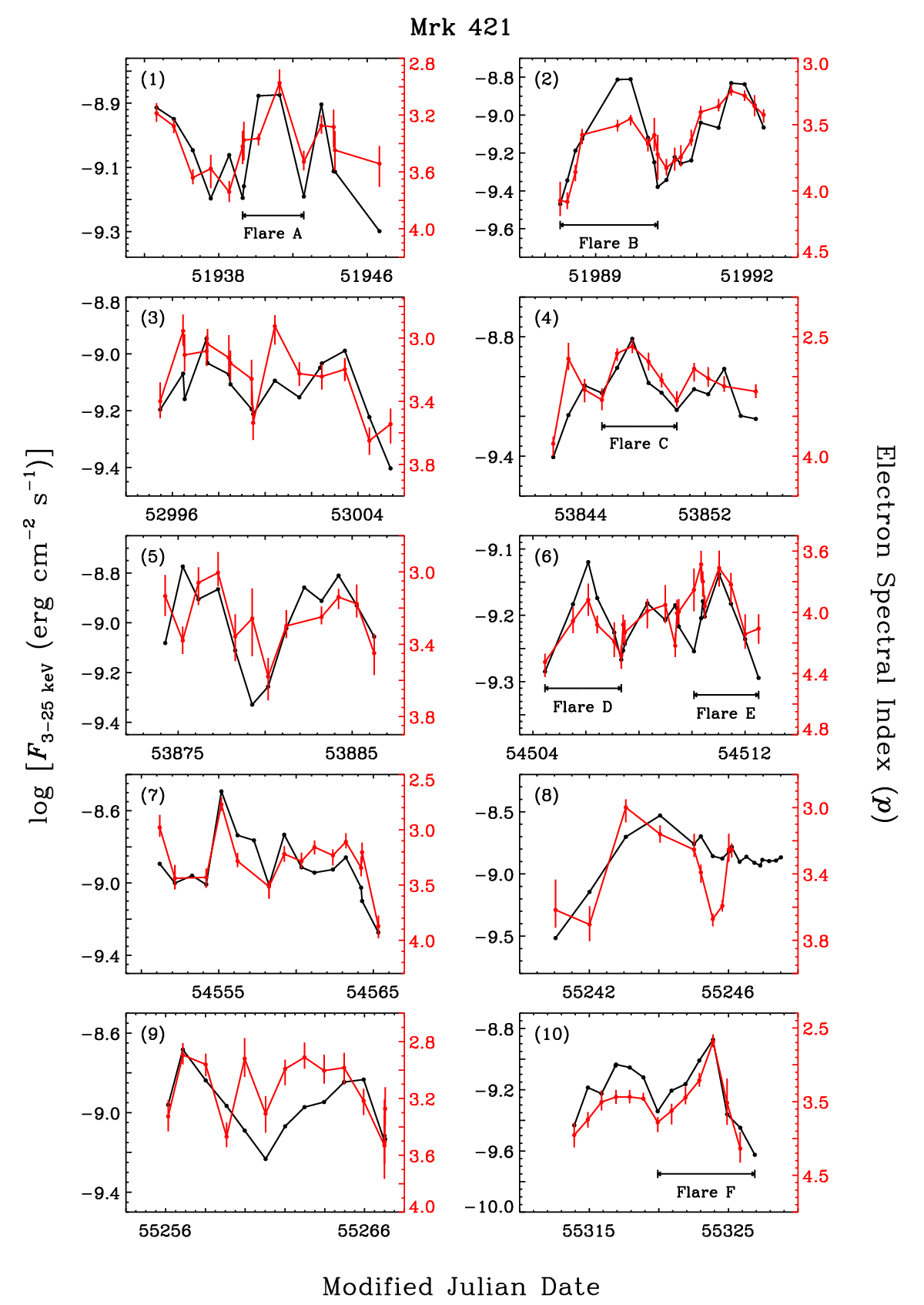}
	\caption{Light curves (black segmented lines; left \textit{y}-axis) and evolution of electron spectral index (red segmented lines; right \textit{y}-axis, which is in the descending order) over time during some typical flares for the five sources. The horizontal segments mark Flares A--L that are further examined in Figure \ref{fig:pfarr}. }
	\label{fig:pfevo}
\end{figure*}

\renewcommand{\thefigure}{\arabic{figure} (Cont.)}
\addtocounter{figure}{-1}

\begin{figure*}[!thbp]
	\centering
	\includegraphics[width=0.76\paperwidth, clip]{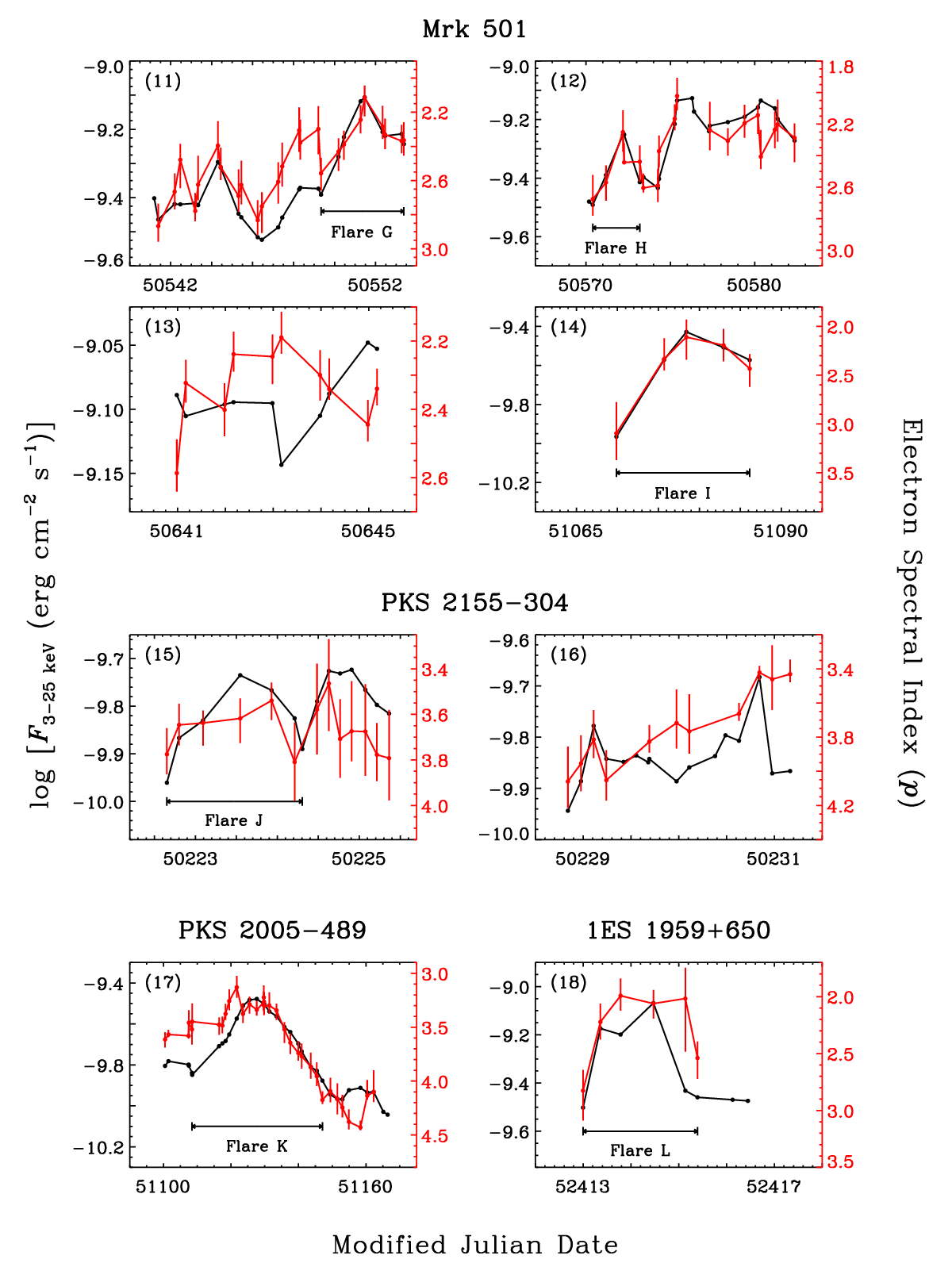}
	\caption{}
\end{figure*}

\renewcommand{\thefigure}{\arabic{figure}}

\begin{figure*}[!thbp]
	\centering
	\includegraphics[width=0.76\paperwidth, clip]{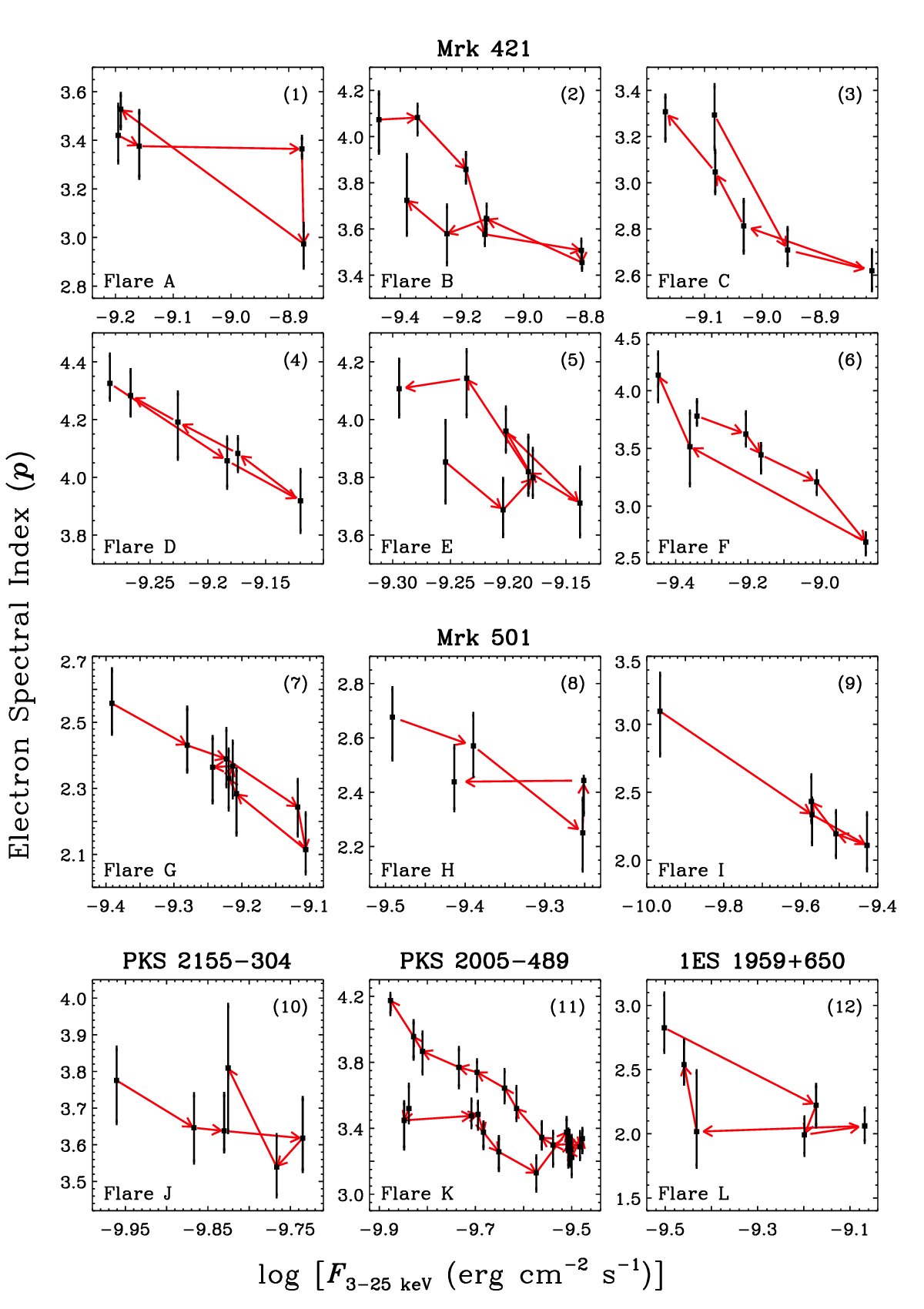}
	\caption{Evolution of electron spectral index ($p$) with flux for the five sources during Flares A--L as annotated in Figure \ref{fig:pfevo}, with red arrows indicating time sequences.}
	\label{fig:pfarr}
\end{figure*}

\section{RESULTS}
\label{sec:results}

\subsection{X-ray Spectra during Flares}
\label{sec:pha}

Figure \ref{fig:spectrum} presents one typical flare and its corresponding X-ray spectra during the flare for each of the aforementioned five sources. It shows that the synchrotron radiation model can describe the spectra very well (see the fitting results in Table~\ref{table:allta}). It is apparent that the X-ray spectrum varies significantly during flares and is harder when flux becomes higher (i.e., harder when brighter), which has been widely studied before: Mrk~421 \citep[e.g.,][]{2000ApJ...541..166F, 2003A&A...402..929B, 2004A&A...424..841R, 2008ApJ...677..906F, 2011ApJ...738...25A, 2013EPJWC..6104013B, 2014A&A...570A..77P, 2015A&A...576A.126A, 2017ApJ...848..103K}, Mrk~501 \citep[e.g.,][]{1998ApJ...492L..17P, 2000A&A...353...97K, 2005ApJ...622..160X, 2006ApJ...646...61G, 2009ApJ...705.1624A, 2017MNRAS.469.1655K}, 1ES~1959+650 \citep[e.g.,][]{2002ApJ...571..763G, 2016MNRAS.457..704K, 2018MNRAS.473.2542K}, PKS~2155--304 \citep[e.g.,][]{2006ApJ...651..782Z, 2006ApJ...637..699Z, 2014MNRAS.444.1077K, 2016NewA...44...21B} and PKS~2005--489 \citep[e.g.,][]{1999ApJ...523L..11P}. 

In the observational energy band (i.e., 3--25 keV shown in Figure~\ref{fig:spectrum}), the spectral shape is different for the five sources, which indicates that the synchrotron radiation peak is located at different energies. Combining the 3--25~keV spectral shape information and the synchrotron peak energy obtained by fitting all PCA spectra with the log-parabolic model detailed in Section~\ref{sec:peakes}, we found that: for Mrk~421, the peak energy of all the spectra is below 6 keV, which is consistent with the results in \citet{2004A&A...413..489M}, \citet{2004ApJ...601..759T} and \citet{2007A&A...466..521T} (but \citealt{2009A&A...501..879T} shows that its peak energy could be up to 30 keV); for Mrk~501, the peak energy of most spectra is above 3 keV, and \citet{2008A&A...478..395M} shows that its peak energy could be up to 100 keV; for PKS~2005--489 and PKS~2155--304, the peak energy of spectra is below 3 keV; for 1ES~1959+650, the peak energy of most spectra is below 30 keV. In fact, it was sometimes difficult to evaluate the exact location of SED peak, which could fall beyond our limited spectral band coverage. Therefore, we could only provide a rough range of peak energy here.

\subsection{Electron Spectral Evolution}
\label{sec:ese}

As we mentioned before, a general trend, which the spectrum hardens with the flux increasing, has been observed in blazars in X-ray observations. There are several conjectures for leading to such a trend. One of them is hardening or softening in the electron spectral distribution. \citet{2006ApJ...647..194X} had demonstrated that variation of electron spectral index ($p$) is indispensable during a flare. They found that the quality of {\it RXTE}/PCA spectra enables utilizing the synchrotron model to place reasonable constraints upon $p$ evolution during major flares of two TeV blazars Mrk~421 and Mrk~501, i.e., $p$ variation is required and the electron spectrum tends to be harder/softer with the increase/decrease of flux, in addition to accompanying changes of some other key parameters. We confirm and strengthen the results of \citet{2006ApJ...647..194X}, by finding that such a trend of $p$ evolution widely exits in multiple flares of five TeV blazars (see Figures~\ref{fig:spectrum} and \ref{fig:pfevo}) and variation of $p$ over time is synchronous with variation of flux over time. In addtion, the trend of $p$ evolution is consistent with that of the $\alpha$ evolution (see Figure \ref{fig:alfacc} in the appendix). Noting the fact that the above five TeV blazars are all HBLs, we further examined the behaviours of BL Lacertae (the brightest IBL in Table \ref{table:allsources}) and 3C 279 (the brightest FSRQ in Table \ref{table:allsources}) in the $\alpha$-flux plot and found that both of them also show a harder-when-brighter trend.

However, there are a few exceptions that show a complex or even opposite evolution of $p$ rather than the simple harder-when-brighter trend during flares. For example, in panels (6), (9), (13), and (16) of Figure~\ref{fig:pfevo}, the count-rate light curve and $p$ ``light curve'' somehow lose track of each other, and thus do not follow the general trend seen in the other panels where $p$ evolution and count-rate evolution generally track each other in a synchronous way. We show the spectra and spectral variations of the case that exhibits the most apparent exception (i.e., panel 13) in Figure \ref{fig:compsed}. These exceptions might be due to the complexity of physical conditions in the emission region and/or the interaction between multiple populations of emitting electrons in two adjacent and comparative flares. For the latter case, the one-zone synchrotron radiation scenario could not be valid and the introduction of multiple populations of emitting electrons might be essential. 

\begin{figure*}[!tbp]
	\centering
	\includegraphics[width=0.85\paperwidth, clip]{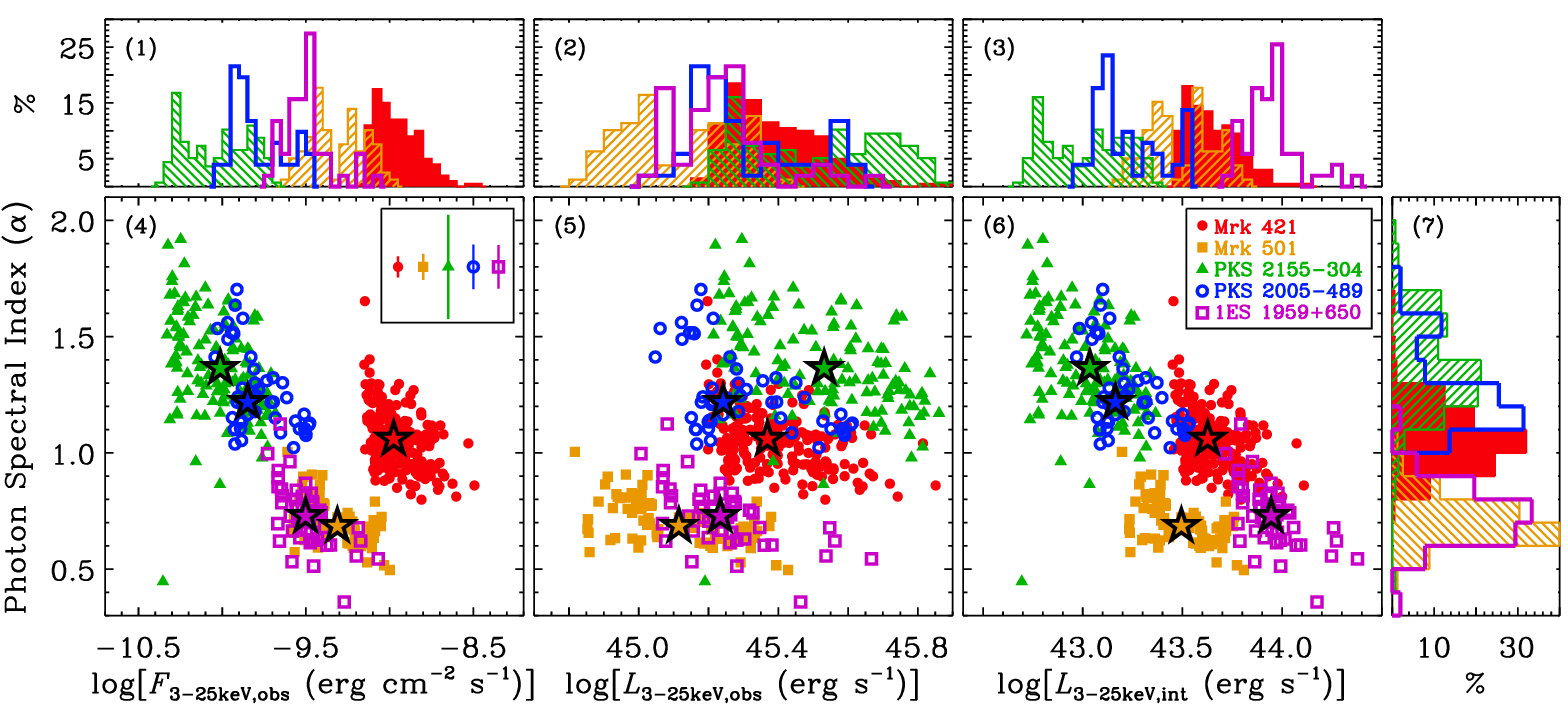}
	\caption{Relations between photon spectral index ($\alpha$) and brightness parameters: observed 3--25 keV X-ray flux ($\textit{F}_{\textrm{3--25keV,obs}}$; panel 4), observed luminosity ($\textit{L}_{\textrm{3--25keV,obs}}$; panel 5) and intrinsic luminosity ($\textit{L}_{\textrm{3--25keV,int}}$ that has been corrected for the Doppler boosting effect using the lower limit of the Doppler factor from \citealt{2013RAA....13..259F}; panel 6) for the five sources, respectively. Pentagrams represent median values. The typical errors on $\alpha$ are shown in the inset at the top-right corner of panel (4). The panels of 1, 2, 3 and 7 show histograms of $\textit{F}_{\textrm{3--25keV,obs}}$, $\textit{L}_{\textrm{3--25keV,obs}}$ and  $\textit{L}_{\textrm{3--25keV,int}}$, $\alpha$, respectively.}  
	\label{fig:alphalumflux}
\end{figure*}

\subsection{Electron Spectral Hysteresis}
\label{sec:esf} 

In a conventional hardness-flux plot, spectral hardness can be different in the rising and falling periods of flares, which is known as ``spectral hysteresis'' and related to both acceleration and cooling timescales. In fact, spectral hysteresis could reveal itself as a ``loop'' shape in the hardness-flux plot (where the spectrum becomes harder along the positive \textit{y}-axis direction and the flux becomes higher along the positive \textit{x}-axis direction). Generally, a ``hard lag'' should result in a counter-clockwise loop, while a ``soft lag'' would lead to a clockwise loop \citep[e.g.,][]{2017ApJ...834....2A}. X-ray spectral hysteresis has been found in many blazars \citep[e.g.,][]{2000ApJ...528..243K, 2002ApJ...581..127B, 2002ApJ...572..762Z, 2002ApJ...574..634S, 2004ApJ...605..662C, 2004A&A...424..841R, 2004ApJ...609..576B, 2005ApJ...622..160X, 2005A&A...443..397B, 2006ApJ...646...61G, 2009ApJ...703..169A, 2009A&A...501..879T, 2016ApJ...831..102K, 2017ApJ...848..103K, 2017MNRAS.469.1655K, 2017Ap&SS.362..196K, 2017ApJ...834....2A, 2018MNRAS.473.2542K}; and UV-optical spectral hysteresis has also been seen in non-blazar jet knots \citep[e.g.,][]{2011ApJ...743..119P}.

From Figure \ref{fig:pfevo}, we further selected a number of flares (i.e., Flares A--L) to examine a different version of spectral hysteresis that is in the form of electron spectral hysteresis (shown as Figure \ref{fig:pfarr}), which is consistent with the photon spectral hysteresis (see Figure \ref{fig:alfaarow} in the appendix). In many of these $p$-flux plots, electron spectral hysteresis is apparent, rendering itself in a ``loop'' (e.g., panel 6 of Figure~\ref{fig:pfarr}  that corresponds to Flare F of Mrk~421), or oblique ``8'' (e.g., panels of 1, 2, 3, 5, 11, and 12 that correspond to Flares A, B, C, E of Mrk~421, Flare K of PKS~2005--489, and Flare L of 1ES~1959+650, respectively) shape; whereas some cases show no apparent hysteresis (e.g.,  panels of 4, 7, 8, 9, and 10 that correspond to Flare D of Mrk~421, Flare G, H, I of Mrk~501, and Flare J of PKS~2155--304, respectively), given the relatively large errors of $p$. As in Sections~\ref{sec:pha} and \ref{sec:ese}, most panels of Figures~\ref{fig:pfarr} also show an overall trend that electron spectrum typically hardens with flux increasing and softens during decreasing phase, which might reflect a process of electron acceleration, injection, or cooling. Interestingly, there are a few cases that seem to behave in a perplexing way. For instance, for Flare A of Mrk~421 (panel~1), the spectrum starts with almost no spectral variability but a flux increase, then suddenly hardens when flux remains invariable (the flare peak might happen to be between the third and fourth observations of this flare, which was not observed and led to such a case) and softens with a flux decrease; and Flare K of PKS~2005--489 (panel~11) shows that the value of $p$ is nearly invariable during the rising period of the flare, given the uncertainties on $p$. \citet{1999ApJ...523L..11P} had analyzed the prominent flare of PKS~2005--489 (as shown in panel 10 of Figure \ref{fig:pfarr}) and found that the 2--10 keV X-ray spectral variability follows a counter-clockwise ``loop'' in the spectral index-flux plane. The evolution of $p$-flux in this work is consistent with their result.

Electron spectral index ($p$) represents the fraction of electrons in different energies. The fraction of high-energy electrons increases with the value of $p$ decreasing. For most flares in Figure~\ref{fig:pfarr}, it appears that the value of $p$ in the rise phase of the flare is larger than (panels 2, 3, 6, 7, 8, and 12) or approximately equal to (panels 4 and 9) the value in the decay phase. In other words, at the beginning of the flare, the fraction of high-energy electrons is low and subsequently increases gradually, which leads to ``hard lag''. However, for the flares in panels (1), (10), and (11), the trend is opposite; this means that the fraction of high-energy electrons is high in the beginning and then decreases, which leads to ``soft lag''. We then used cross-correlation function to search for likely time lags between soft-band (3--8 keV) and hard-band (8--25 keV) light curves, but found no evidence for existence of time lags.
These non-detections of time lags might be due to the fact that the actual time lag is likely intra-day, which is difficult to resolve using our light curves of several-day time resolution. In \citet{2010ApJ...722..358G}, two flares of Mrk~421 lasting for 0.5 days showed several-hour time lags between the 0.5--2~keV and 2--10~keV light curves, and displayed different movements in the hardness-flux plot. \citet{2000A&A...353...97K} indicated that the time lag between 3~keV and 25~keV is smaller than 15 hours in Mrk~501. Therefore, more intra-day observational data would be required to detect likely time lags of our sources that might be responsible for the observed electron spectral hysteresis.

\begin{figure}[!tbp]
	\centering
	\includegraphics[width=\linewidth, clip]{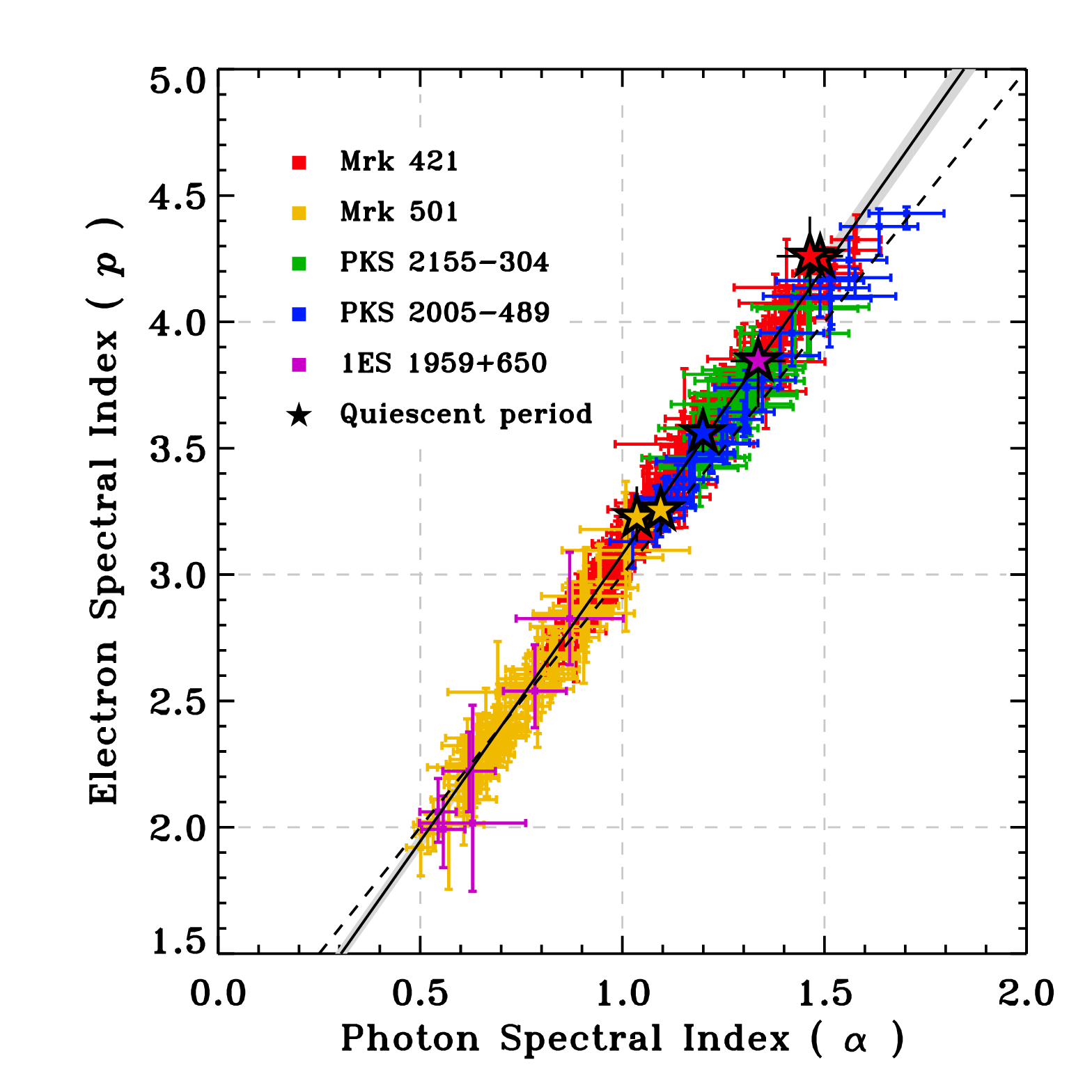}
	\caption{$p$-$\alpha$ (electron spectral index-photon spectral index) relation of the five sources. Colored filled squares represent flaring observations; pentagrams represent quiescent-period observations (see Section~\ref{sec:ldata} and Table~\ref{table:allta}). Squares and pentagrams in the same color represent the same source. The black solid line and its shaded region represent the best fit and 1-$\sigma$ uncertainty to all data of flaring periods, i.e., $p$ = ($2.27 \pm 0.03)\times \alpha$ + ($0.81 \pm 0.03$). The dashed line indicates the relation of $\alpha=(p-1)/2$ as expected in the optically thin synchrotron radiation spectrum in power-law shape.} 
	\label{fig:pa}
\end{figure}

\begin{figure}[!tbp]
	\centering
	\includegraphics[width=\linewidth, clip]{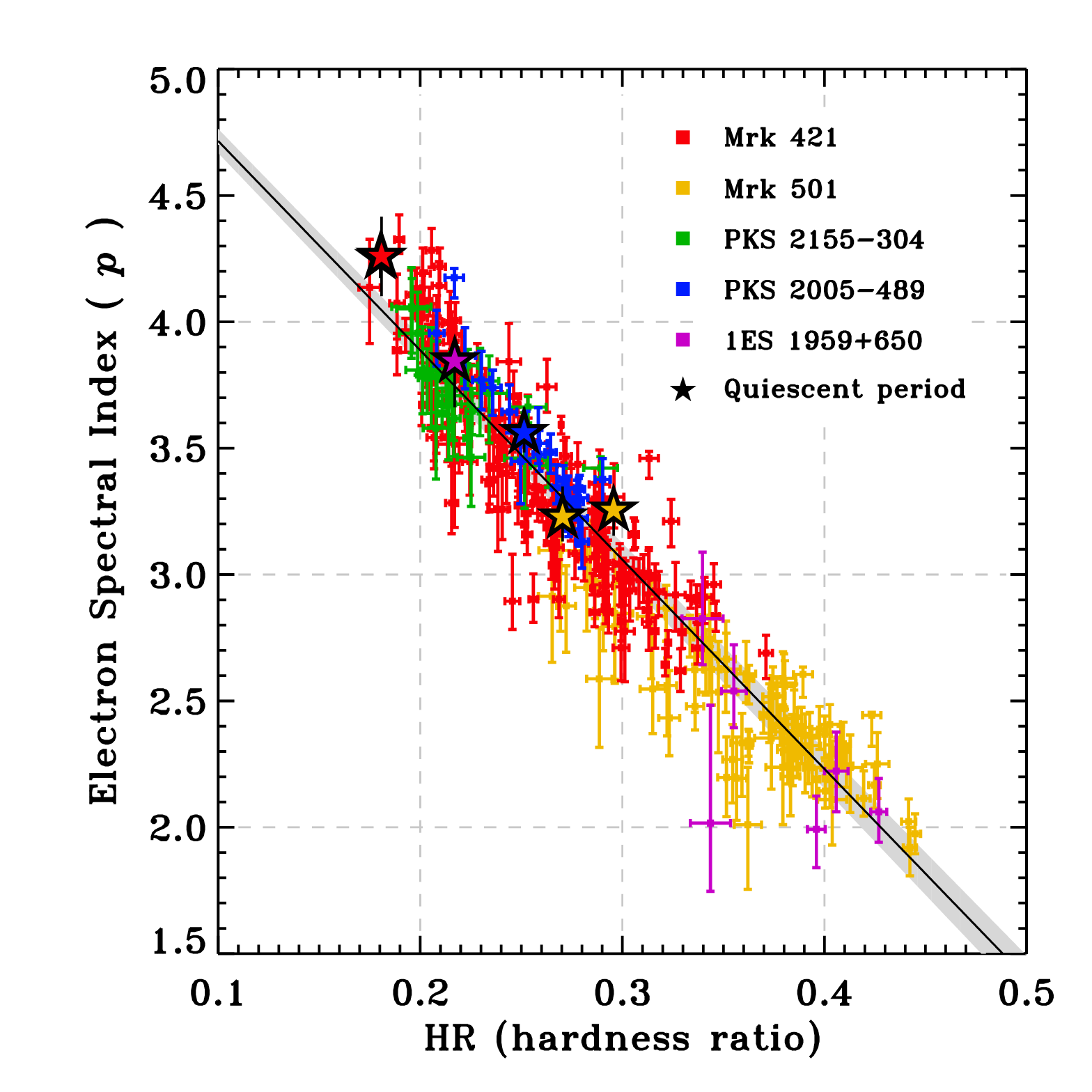}
	\caption{$p$-HR (electron spectral index-hardness ratio) relation of the five sources. All symbols have the same meaning as in Figure~\ref{fig:pa}. The best fit to all data of flaring periods is $p$ = ($-8.28 \pm 0.16)\times$HR + ($5.54 \pm 0.05$).}
	\label{fig:phr}
\end{figure}

\subsection{Photon Spectral Index versus Luminosity}
\label{sec:alphalum}

As Figure \ref{fig:alphalumflux} shows, for a single object, photon spectral index ($\alpha$) decreases with increasing flux (panel 4) or luminosity (panels of 5, 6), which is the so-called harder-when-brighter trend (see Section~\ref{sec:pha}); while for the entirety of the five sources, photon spectral index seems to increase with increasing observed luminosity (panel 5; see median values denoted by pentagrams), but decrease with increasing intrinsic luminosity that has been corrected for the Doppler boosting effect (panel 6). To control for spectral data quality in Figure~\ref{fig:alphalumflux}, we only chose observations with fluxes above 0.25$\times$($\textit{F}_\textrm{max}-\textit{F}_\textrm{min}$) for each source, where $\textit{F}_\textrm{max}$ and $\textit{F}_\textrm{min}$ are the maximum and minimum fluxes of the source among the 16-year data, respectively.

The positive correlation between photon spectral index and observed luminosity (panel 5 of Figure \ref{fig:alphalumflux}) is correlated with the ``blazar sequence'' \citep{1998MNRAS.299..433F,1998MNRAS.301..451G}. According to the blazar sequence, there is a negative correlation between the synchrotron peak energy and the observed bolometric luminosity ($L_{\rm bol}$). The observed bolometric flux can be estimated roughly through the relation of $\textit{F}_\textrm{bol} \simeq \textrm{5} \nu_\textrm{p} \textit{F}\textrm{(}\nu_\textrm{p}\textrm{)}$, where $\nu_\textrm{p} \textit{F}\textrm{(}\nu_\textrm{p}\textrm{)}$ is the peak flux of the synchrotron hump \citep{2004A&A...413..489M}, such that $L_{\rm bol}\simeq 5 \nu_\textrm{p} L(\nu_\textrm{p})\propto L_\textrm{3--25 keV}$. In addition, there is an anti-correlation between X-ray photon spectral index and synchrotron peak energy \citep[e.g.,][]{1999ApJ...525..191L,2005A&A...434..385G,2005ApJ...625..727P}. Therefore, when the peak moves to the lower energy band, the X-ray luminosity will increase and the X-ray spectrum will tend to steepen. As a result, there would be a positive correlation between the X-ray luminosity and X-ray photon spectral index, as demonstrated in the panel (5) of Figure~\ref{fig:alphalumflux} (the Spearman's ranking correlation for the overall $\alpha$-$L_\textrm{3--25keV,observed}$ relation of the five sources is 0.31 with a significance of 3.42 $\times \textrm{10}^\textrm{--13}$). However, the anti-correlation between synchrotron peak energy and luminosity (the blazar sequence) dispears after applying Doppler boosting correction to the observed luminosity \citep[e.g.,][]{2008A&A...488..867N,2009RAA.....9..168W,2014JApA...35..381H, 2017ApJ...835L..38F}. In this work, we performed approximate Doppler-corrections using the equation $\textit{L}_\textrm{intrinsic} = \textit{L}_\textrm{observed}/\delta^3$, where $\textit{L}_\textrm{intrinsic}$ is the intrinsic luminosity, $\textit{L}_\textrm{observed}$ is the observed luminosity and $\delta$ is the lower limit of the $\gamma$-ray Doppler factor from \citet{2013RAA....13..259F} that is estimated according to the pair-production optical depth \citep{1993ApJ...410..609M}: 2.77 for Mrk~421, 2.83 for Mrk~501, 4.15 for PKS~2155--304, 3.30 for PKS~2005--489 and 2.32 for 1ES~1959+650. Our results are consistent with the previous studies (panel 6 of Figure \ref{fig:alphalumflux}): the Spearman's ranking correlation for the overall $\alpha$-$L_\textrm{3--25keV,intrinsic}$ relation of the five sources is --0.46 with a significance of 2.33$\times \textrm{10}^\textrm{--28}$.

\section{DISCUSSION}
\label{sec:discussion}

\begin{figure*}[!thbp]
	\centering
	\includegraphics[width=0.85\paperwidth, clip]{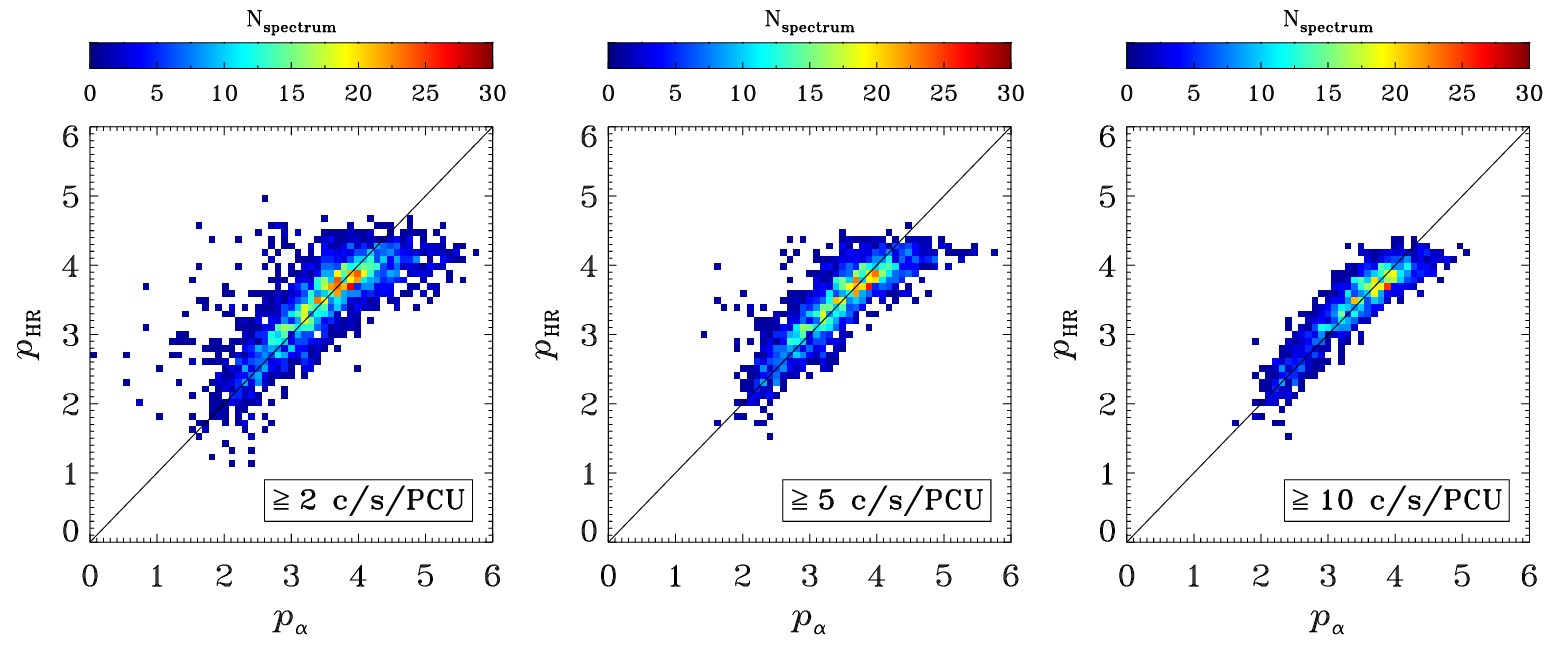}
	\caption{Density maps of $p_{\rm HR}$-$p_\alpha$ distribution ($p_{\rm HR}$ and $p_\alpha$ estimated using the $p$-HR and $p$-$\alpha$ relations, respectively) for all HBL spectra in three different count-rate regimes, respectively. Black solid lines represent $y=x$. Color bars atop indicate numbers of HBL spectra contained within a pixel.}
	\label{fig:pesti}
\end{figure*}

\subsection{Electron Spectral Index versus Photon Spectral Index}
\label{sec:esipsi}
We obtained the photon spectral index (\textrm{$\alpha$}) by fitting spectra with the cut-off power law model rather than the log-parabolic model. These two models both provided great spectral fitting results, but the photon spectral index in the log-parabolic model is energy dependent, so we did not adopt $\alpha$ from this model. The cut-off power law model follows the relation of $F(E) \propto E^{-\alpha}\cdot\exp{(-E/\beta)}$, where $\beta$ is the e-folding energy of exponential roll-off. In this work, we have assumed that the electron spectral distribution follows the power-law shape, which usually produces the optically thin synchrotron radiation spectrum with a power-law photon spectral index of $\alpha$ = $(p-\textrm{1})/\textrm{2}$ (see the dashed line in Figure~\ref{fig:pa}). However, in this work, the cut-off power-law model could provide better fits to all the spectra than the power-law model; therefore, we used the cut-off power law model to obtain photon spectral index ($\alpha$). 
   
According to Figure \ref{fig:pa}, there is a significant linear relation between $p$ and $\alpha$ during flares of the five sources, which follows the formula of 
\begin{equation}
p =(2.27 \pm 0.03)\times \alpha + (0.81 \pm 0.03).
\label{equ:pa}
\end{equation}
This relation shows a slight deviation from the theoretical relation of $\alpha$ = $(p-\textrm{1})/\textrm{2}$ for the power-law spectral distribution. This deviation might be due to the energy loss of electrons and acceleration process of relativistic electrons, which could produce the spectrum not exactly following the power-law distribution. Therefore, for the X-ray spectrum not following the power-law shape, it might not be suitable to use $\alpha$ = $(p-\textrm{1})/\textrm{2}$ to calculate $p$ using $\alpha$ and vice versa.

In addition, values of $p$ and $\alpha$ in relatively quiescent periods (i.e., pentagrams in Figure~\ref{fig:pa}) seem to follow nicely the above $p$-$\alpha$ relation that was derived with flaring periods, which indicates that the spectra during flaring and quiescent periods share the same $p$-$\alpha$ relation.

\subsection{Electron Spectral Index versus Hardness Ratio}
\label{sec:esihr}

We define hardness ratio (HR) as HR = \textit{H}$/$\textit{S}, where \textit{H} and \textit{S} are count rates in the 8--25 keV and 3--8 keV bands, respectively. As expected, Figure~\ref{fig:phr} also presents a linear relation between $p$ and HR during flares, which follows the formula of 
\begin{equation}
p = (-8.28 \pm 0.16)\times \textrm{HR} + (5.54 \pm 0.05). 
\label{equ:phr}
\end{equation}
We had verified that a similar linear relation between $p$ and HR$_{(\textrm{10}\textrm{-}\textrm{25}) \textrm{keV}/(\textrm{3}\textrm{-}\textrm{10}) \textrm{keV}}$ would also be obtained, which is not presented here. The $p$-HR relation (i.e., Eq.~\ref{equ:phr}) is not as tight as the $p$-$\alpha$ relation (i.e., Eq.~\ref{equ:pa}), and the former has significantly larger scatters than the latter, which is also expected given the following two facts: the derivation of $\alpha$ makes use of full spectral information while the calculation of HR only utilizes crude spectral information; and the influence of Galactic absorption was not taken into account for deriving HR, which should introduce additional small scatters. Furthermore, values of $p$ and HR in relatively quiescent periods (i.e., pentagrams in Figure~\ref{fig:phr}) also seem to generally follow the $p$-HR relation that was derived with flaring periods (cf.
Section~\ref{sec:esipsi} and Figure~\ref{fig:pa}).

\subsection{Application of $p$-$\alpha$ and $p$-HR Relations}
\label{sec:apply}
The $p$-$\alpha$ and $p$-HR relations provide us two quick and straightforward empirical approaches to roughly estimate the electron spectral index ($p$) simply based on values of $\alpha$ and/or HR, without resorting to detailed synchrotron radiation modeling. The distribution of $p$ can only be obtained through fitting spectra with the synchrotron radiation model, which not only relies on high-quality spectral data, but also takes a long time. Therefore, if one is only interested in knowing the approximate range of $p$ for a particular spectrum, then it would be efficient to estimate $p$ using one of the $p$-$\alpha$ and $p$-HR relations or even both.

For the purpose of verifying the reliability of these two approaches, we compared values of $p$ estimated using $\alpha$ with that estimated using HR,  utilizing all HBL spectral data among the 32 TeV blazars (note that the 3--25 keV spectra of FSRQs, LBLs, and IBLs might be dominated by both synchrotron and inverse Compton scattering components, which are not suitable for simple synchrotron radiation modeling). Figure~\ref{fig:pesti} shows a reasonably good agreement between $p_{\alpha}$ and $p_{\textrm{HR}}$ that were derived using the $p$-$\alpha$ and $p$-HR relations, respectively. As the flux increases, the estimation of $p$ becomes more reliable, leading to an improved agreement between $p_{\alpha}$ and $p_{\textrm{HR}}$, thanks to the better quality of data. Therefore, although the $p$-HR relation shows a larger scatter compared with the $p$-$\alpha$ relation, it is still a reliable way to estimate $p$ quickly.

One thing worth noting is that there is a nearly horizontal low-density tail to the top of the distribution. We excluded the likely ``pileup''-like effect for this feature because the fluxes of these observations are not very high. The possible reason is that $p$-$\alpha$ and $p$-HR relations do not completely follow the linear correlation, i.e., the $p$ derived from the data is smaller than the best-fit $p$-$\alpha$ relation at $\alpha$ $>$ 1.5 (see Figure \ref{fig:pa}) and is larger than the best-fit $p$-HR relation at HR $<$ 0.2 (see Figure \ref{fig:phr}). Therefore, if we use the best-fit relations to estimate the $p$ in the high $p$ range, the $p_{\alpha}$ tends to be larger and the $p_{\textrm{HR}}$ tends to be smaller, which will lead to a nearly horizontal tail. Even so, for the vast majority of observations, $p$-$\alpha$ and $p$-HR relations can provide consistent $p$ estimates.

As demonstrated above, both the $p$-$\alpha$ and $p$-HR relations are suitable for estimation of $p$ in the case that the radiative process is dominated by synchrotron radiation in jets. 
We note that the $p$-HR relation should be instrument-dependent because HR is closely related to the instrument response, therefore being valid only for \textit{RXTE}/PCA data; while the $p$-$\alpha$ relation should be instrument-independent.

\begin{figure}[!tbp]
	\centering
	\includegraphics[width=\linewidth, clip]{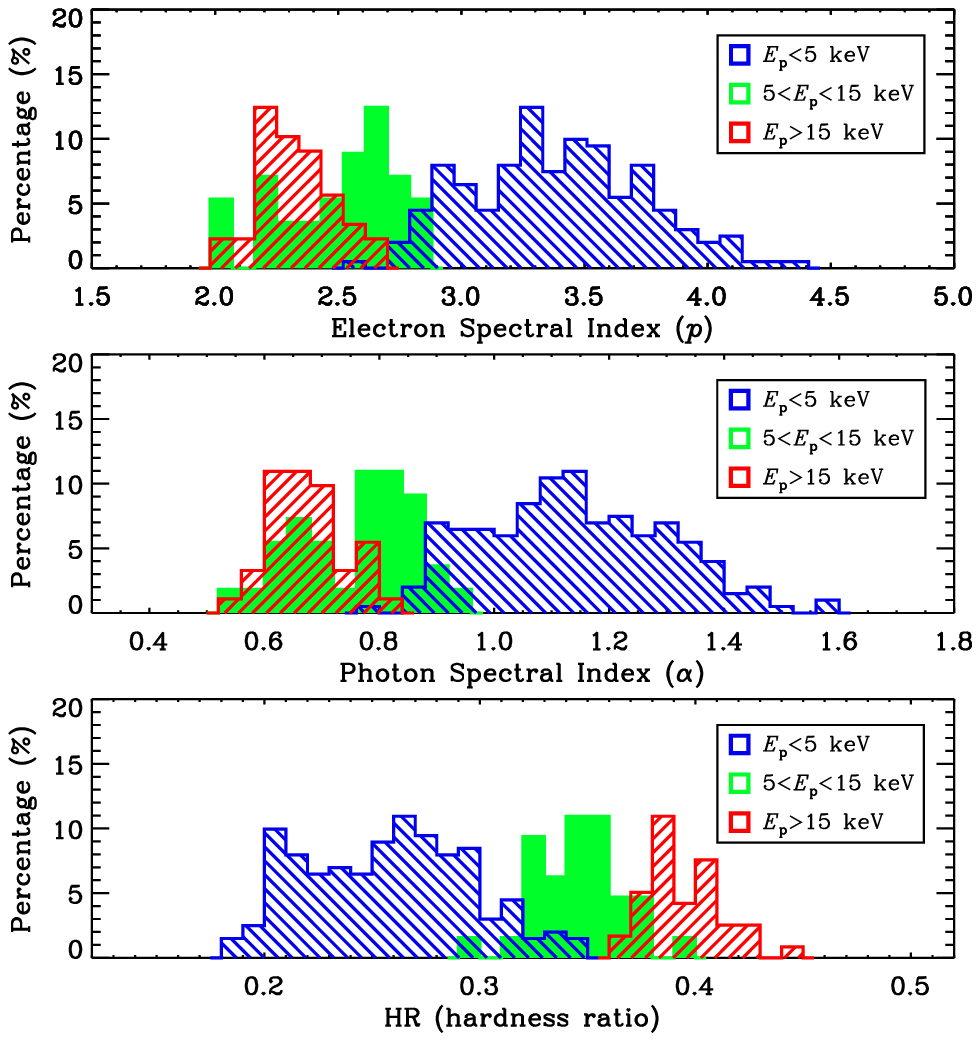}
	\caption{Distributions of electron spectral index, photon spectral index, and hardness ratio
		for all spectra of flaring periods, which are divided into three groups according
		to their peak energies ($E_{\rm p}$) of synchrotron radiation humps
		(blue, green, and red histograms for $E_{\rm p}<5$~keV, 5~keV$<E_{\rm p}<$15~keV, and $E_{\rm p}>$15~keV, respectively).}
	\label{fig:pahr}
\end{figure}

\subsection{Peak Energy versus Spectral Parameters}
\label{sec:peakes}

For many cases in blazars, the log-parabolic model could greatly reproduce the spectra around the synchrotron peak in the SED, which provides a valid method to estimate the energy and flux of the peak \citep[e.g.,][]{2004A&A...413..489M,2004ApJ...601..759T,2007A&A...466..521T,2009A&A...501..879T}. Therefore, we used the log-parabolic model to estimate $\textit{E}_{\textrm{p}}$ following \cite{2004A&A...413..489M}, which is given by
\begin{equation}
\textit{F(E)}=\textit{K}(\textit{E}/\textit{E}_1)^{(-\textit{a}+\textit{b} \log(\textit{E}/\textit{E}_1))}\ \ \bigl(\textrm{ph}\ \textrm{cm}^{\textrm{--2}}\ \textrm{s}^{\textrm{--1}}\ \textrm{keV}^{\textrm{--1}}\bigr).
\label{equ:lppp}
\end{equation}
$\textit{E}_\textrm{1}$ is the reference energy that is generally fixed to 1 keV, \textit{a} is the photon spectral index at the energy of $\textit{E}_\textrm{1}$, \textit{b} is the curvature parameter, and \textit{K} is the normalization factor. The values of these parameters can be derived from the spectral fitting process. The peak energy of synchrotron radiation hump is given by $\textit{E}_{\textrm{p}}^{\textrm{obs}}=\textit{E}_\textrm{1} \textrm{10}^{(\textrm{2}-\textit{a})/\textrm{2}\textit{b}}.$ The rest-frame peak energy is $\textit{E}_{\textrm{p}}=(1+\emph{z})\textit{E}_{\textrm{p}}^{\textrm{obs}}$, where \emph{z} is the redshift. In some cases, parameter \textit{b} is below 0, which means that the fitted curve is concave, and the resulting peak energy is not the real peak energy of synchrotron hump. There are many reasons for such a result, such as a concave spectrum and poor quality of data (especially in the high-energy band). For such cases, we could only obtain a rough range of $\textit{E}_{\textrm{p}}$: if \textit{b} $<$ 0 and $\textit{E}_{\textrm{p,fit}}$ $<$ 3 keV, then $\textit{E}_{\textrm{p}}$ $>$ 25 keV or $\textit{E}_{\textrm{p}}$ $<$ 3 keV; if \textit{b} $<$ 0 and $\textit{E}_{\textrm{p,fit}}$ $>$ 3 keV, then $\textit{E}_{\textrm{p}}$ $<$ 3 keV ($\textit{E}_{\textrm{p,fit}}$ is the fitting result of the peak energy). Fortunately, there are only two observations with \textit{b} $<$ 0 and both of them belong to the second case.  

For the flaring periods of the five sources, according to the location of synchrotron radiation SED peak, we roughly divided a total of 276 spectra into three groups: $\textit{E}_{\textrm{p}}$ $<$ 5 keV (202 spectra), 5 keV $<$ $\textit{E}_{\textrm{p}}$ $<$ 15 keV (32 spectra), and $\textit{E}_{\textrm{p}}$ $>$ 15 keV (42 spectra). Figure~\ref{fig:pahr} presents the distributions of three spectral parameters $p$, $\alpha$, and HR for these three groups of spectra. Although our sample is not complete, it still reveals a general trend that, with the synchrotron radiation SED peak energy increasing, both $p$ and $\alpha$ decrease while HR increases. This is consistent with the trend that the spectra are harder with higher peak frequencies seen in other works \citep[e.g.,][]{1999ApJ...525..191L,2005A&A...434..385G,2005ApJ...625..727P}.

Additionally, some rough constraints upon $\textit{E}_{\textrm{p}}$ (also see Section~\ref{sec:pha}) and the three spectral parameters can be obtained, according to Figure~\ref{fig:pahr}:
if $\textit{E}_{\textrm{p}}$ is lower than 5~keV, then $p$ is higher than $\sim$ 2.7, $\alpha$ is higher than $\sim$ 0.8, and HR is lower than $\sim$ 0.35; if $\textit{E}_{\textrm{p}}$ is higher than 5 keV, then $p$ is lower than $\sim$ 3.0, $\alpha$ is lower than $\sim$ 1.0, and HR is higher than $\sim$ 0.30. The constrained range of $\alpha$ is in good agreement with the result in \citet{2005ApJ...625..727P}, where three hard ($\Gamma$ $<$ 2, i.e., $\alpha$ $<$ 1) spectra correspond to $\textit{E}_{\textrm{p}}$ $>$ 5 keV.

\section{SUMMARY AND CONCLUSIONS}
\label{sec:summary}

During its entire lifespan ($\sim$ 16 years), \textit{RXTE} had observed 32 TeV blazars, including 2 FSRQs, 1 LBLs, 5 IBLs, and 24 HBLs. 
In this paper, we analyzed the 16-year \textit{RXTE}/PCA observational data of the 32 TeV blazars and further selected out five brightest sources 
to carry out a systematic investigation of X-ray spectral variability during their major flares in the \textit{RXTE} era, using both empirical spectral fitting (to obtain values of $\alpha$, flux and $E_{\rm p}$) and theoretical synchrotron radiation modeling (to obtain $p$ distributions).
Our work builds on \citet{2006ApJ...647..194X} that studied only two TeV blazars, confirms and strengthens their main results with a larger sample, and provides 
many further insights regarding X-ray spectral variability of TeV blazars.
We summarize our main results as follows:

\begin{enumerate}

\item 
The cut-off power-law and log-parabolic models could provide evenly good fitting results to the X-ray spectra of all sources. 
The X-ray spectra, characterized by $\alpha$, display a harder-when-brighter trend during a number of flares of the five brightest sources (i.e., Mrk~421, Mrk~501, PKS~2155--304, PKS~2005--489, and 1ES~1959+650), which is consistent with previous studies.  

\item
The high quality of the PCA data of the five sources enables detailed synchrotron radiation modeling upon their spectra.
It seems clear that the evolution of 
$p$ also generally follows a harder-when-brighter trend; 
and the variation of $p$, accompanied by changes of other key parameters, is required to explain the observed X-ray spectral variability of TeV blazars during flares, which would have useful implications for interpreting the associated higher-energy (i.e., gamma-ray) spectral variability that the same population of ultra-relativistic electrons are responsible for.
These results confirm and strengthen that of \citet{2006ApJ...647..194X}.
However, there are some cases that do not follow the harder-when-brighter trend exactly,
which might be related to the complex physical conditions in the emitting region or the ``contamination'' of electron populations from adjacent flares.
 
\item
Electron spectral hysteresis is clearly seen in many but not all $p$-flux plots,
rendering itself in a ``loop'' or oblique ``8'' shape.
Although this phenomenon is often associated with time lags between the soft and hard bands, 
no apparent hard or soft lag is identified based on our several-day-timescale light curves.
Intra-day observations might help resolve likely intra-day time lags.

\item
A tight $p$-HR relation and a tighter $p$-$\alpha$ relation are obtained
using spectra of flaring periods,
both of which are also applicable to stacked data of quiescent periods,
indicating that both relations are independent of flux level.
These two relations can be used to estimate $p$ quickly and straightforwardly,
and the reliability of $p$ estimation improves as improved data quality.

\item
Collectively (i.e., TeV blazars being treated as a whole),
$\alpha$ and X-ray luminosity are positively correlated,
$E_{\rm p}$ is negatively correlated with $p$ and $\alpha$,
and $E_{\rm p}$ is positively correlated with HR.
All these correlations are in line with the blazar sequence.
However, after correcting for the Doppler boosting effect, 
$\alpha$ and intrinsic X-ray luminosity follow an anti-correlation.

\end{enumerate}

\acknowledgments

We thank the referees for helpful comments that improved this paper. Y.J.W. and Y.Q.X. acknowledge support from the 973 Program (2015CB857004), the National Natural Science Foundation of China (NSFC-11473026, 11421303), and the CAS Frontier Science Key Research Program (QYZDJ-SSW-SLH006). J.H.F acknowledges support from the grant of NSFC-11733001.

\appendix

\section{photon spectral evolution and photon spectral hysteresis}
\label{alfaevo}

In Figures \ref{fig:pfevo} and \ref{fig:pfarr}, we have shown the electron spectral evolution and electron spectral hysteresis, respectively. Given that most of the previous works studied spectral variability through photon spectral index, in this appendix we also show the photon spectral evolution in Figure \ref{fig:alfacc} and photon spectral hysteresis in Figure \ref{fig:alfaarow}. The photon spectral evolution shows a harder-when-brighter trend (see Figure \ref{fig:alfacc}), which is consistent with previous studies, and also shows a similar trend to that revealed by $p$ evolution (see Figure \ref{fig:pfevo}). In addition, the photon spectral hysteresis (see Figure \ref{fig:alfaarow}) shows a similar trend to that seen with electron spectral hysteresis (see Figure \ref{fig:pfarr}). Therefore, consistent results on spectral evolution and hysteresis are obtained using either the $\alpha$ or $p$ representation, which is expected given the tight $p$--$\alpha$ relation (see Figure \ref{fig:pa}).

\section{opposite evolution of spectral index with flux}
\label{comflare}
During the period between MJD 50641 and 50645 (i.e., panel 13 of Figure \ref{fig:pfevo}), Mrk~501 shows a softer-when-brighter trend in terms of $p$ variation, which is opposite to the harder-when-brighter trend existing in most of the studied cases (see Figure \ref{fig:pfevo} and Section \ref{sec:ese}). The spectra and $\alpha$ variations of these observations are showed in Figure \ref{fig:compsed}, which also show a softer-when-brighter trend. Several cases also present a similar opposite trend, e.g., Mrk~421 in the period between MJD 54509 and 54511 (see panel 6), and PKS~2155--304 in the period between MJD 50229.2 and 50230.2 (see panel 16). This opposite trend of these three cases exists in the transition region between two individual flares, which indicates that it might be due to the interaction between multiple populations of emitting electrons in the two adjacent flares.   \\

\bibliographystyle{apj_hyperref}
\bibliography{ms_new}

\setcounter{figure}{0}
\renewcommand{\thefigure}{A\arabic{figure}}

\begin{figure*}[!thbp]
	\centering
	\includegraphics[width=0.8\paperwidth, clip]{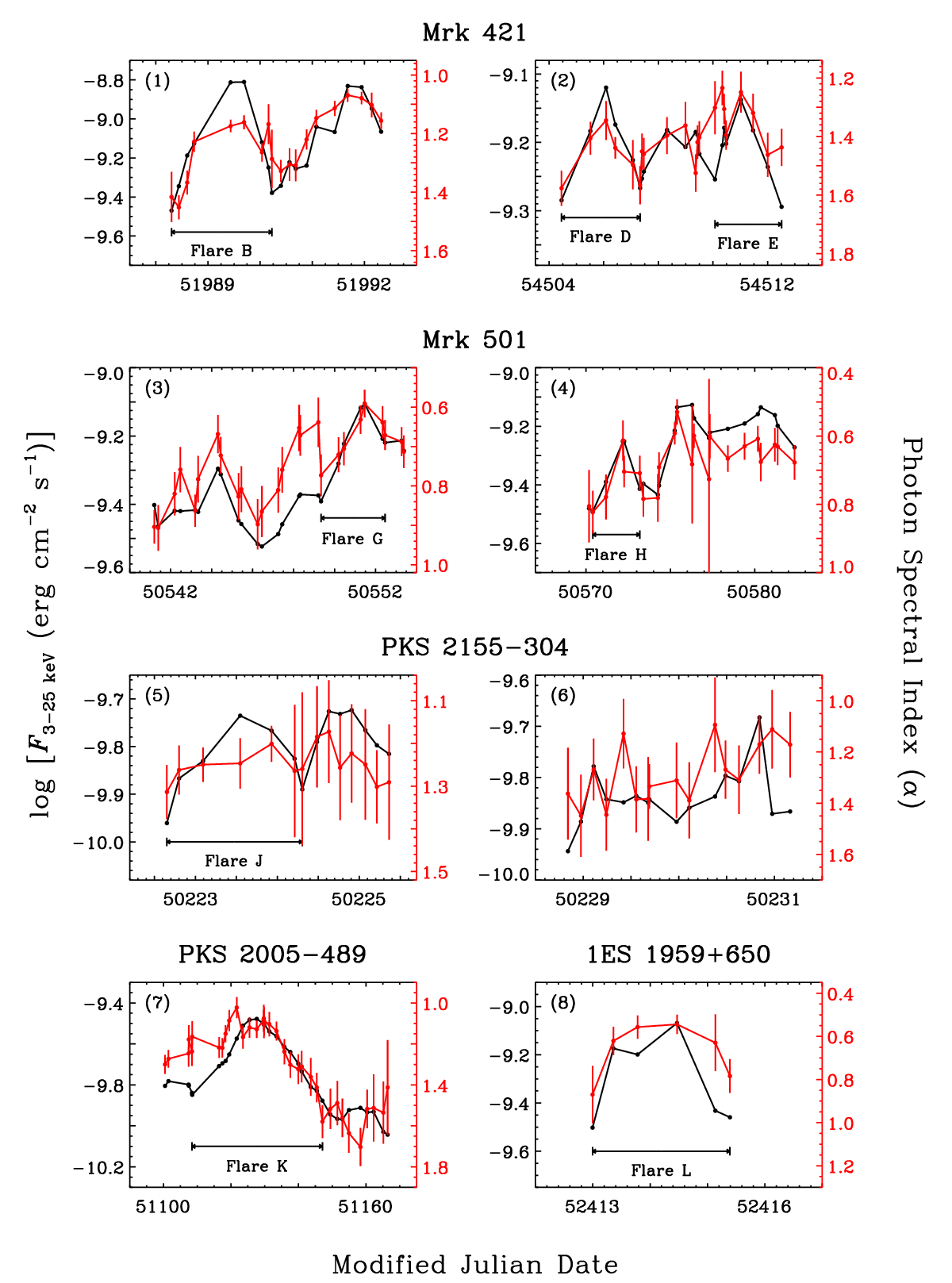}
	\caption{Same as Figure \ref{fig:pfevo}, but for the evolution of photon spectral index ($\alpha$; red segmented lines; right y-axis). The $\alpha$ evolution shows a harder-when-brighter trend, similar to the $p$ evolution. For brevity, only 8 panels of Figure 2 are re-plotted here for demonstration, as all panels give the same result.}
	\label{fig:alfacc}
\end{figure*}

\begin{figure*}[!thbp]
	\centering
	\includegraphics[width=0.76\paperwidth, clip]{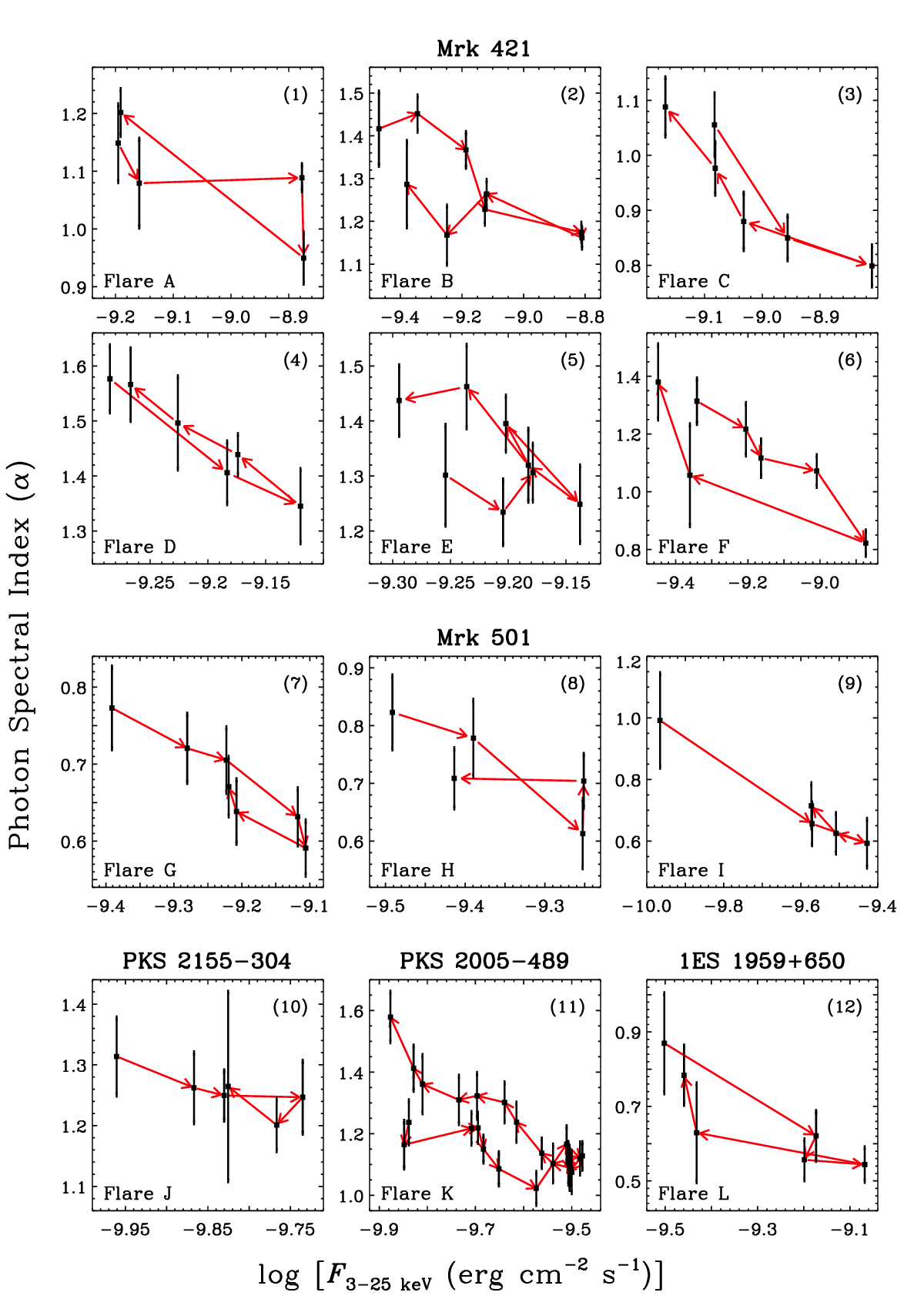}
	\caption{Same as Figure \ref{fig:pfarr}, but for the evolution of photon spectral index ($\alpha$).}
	\label{fig:alfaarow}
\end{figure*}

\setcounter{figure}{0}
\renewcommand{\thefigure}{B\arabic{figure}}

\begin{figure*}[!thbp]
	\centering
	\includegraphics[width=0.32\paperwidth, clip]{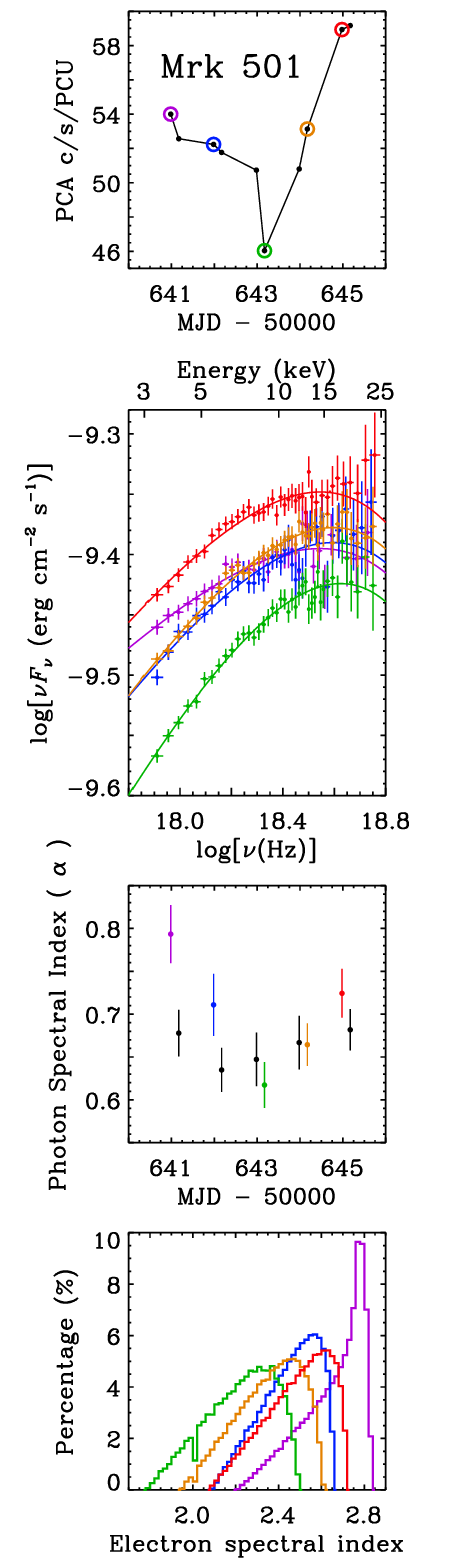}
	\caption{ Same as Figure \ref{fig:spectrum}, but for Mrk~501 during the period between MJD 50641 and 50645 (cf. panel 13 of Figure \ref{fig:pfevo}).} 
	\label{fig:compsed}
\end{figure*}

\end{document}